\documentstyle[preprint,aps,psfig,multicol]{revtex}
\tightenlines

\newcommand{\Done}{\bbox{\Delta}^{(\pm)}}
\newcommand{\Dtwo}{\bbox{\Delta}^{(2)}}
\newcommand{\LmQCD}{\Lambda_{\mbox{\scriptsize QCD}}}

\begin{document}
\draft
\preprint{\vbox{\hbox{HUPD-9904}}}

\title{$B$ meson leptonic decay constant with quenched lattice NRQCD}
\author{(JLQCD Collaboration) \\
  K-I.~Ishikawa$^{\mbox{\scriptsize\ a, b}}$,
  N.~Yamada$^{\mbox{\scriptsize\ a}}$,
  S.~Aoki$^{\mbox{\scriptsize\ c}}$,
  M.~Fukugita$^{\mbox{\scriptsize\ d}}$,
  S.~Hashimoto$^{\mbox{\scriptsize\ b}}$,
  N.~Ishizuka$^{\mbox{\scriptsize\ c, e}}$,
  Y.~Iwasaki$^{\mbox{\scriptsize\ c, e}}$,
  K.~Kanaya$^{\mbox{\scriptsize\ c, e}}$,
  T.~Kaneda$^{\mbox{\scriptsize\ c}}$,
  S.~Kaya$^{\mbox{\scriptsize\ b}}$,
  Y.~Kuramashi$^{\mbox{\scriptsize\ b, f}}$,
  H.~Matsufuru$^{\mbox{\scriptsize\ a}}$,
  M.~Okawa$^{\mbox{\scriptsize\ b}}$,
  T.~Onogi$^{\mbox{\scriptsize\ a}}$,
  S.~Tominaga$^{\mbox{\scriptsize\ b}}$,
  N.~Tsutsui$^{\mbox{\scriptsize\ a}}$,
  A.~Ukawa$^{\mbox{\scriptsize\ c, e}}$,
  T.~Yoshi\'e$^{\mbox{\scriptsize\ c, e}}$
}

\address{ 
  $^{\mbox{\scriptsize\ a}}$
  Department of Physics, Hiroshima University, 
  Higashi-Hiroshima, Hiroshima 739-8526, Japan\\ 
  $^{\mbox{\scriptsize\ b}}$
  High Energy Accelerator Research Organization (KEK),
  Tsukuba, Ibaraki 305-0801, Japan\\ 
  $^{\mbox{\scriptsize\ c}}$
  Institute of Physics, University of Tsukuba, 
  Tsukuba, Ibaraki 305-8571, Japan\\ 
  $^{\mbox{\scriptsize\ d}}$
  Institute for Cosmic Ray Research, 
  University of Tokyo, 
  Tanashi, Tokyo 188-8502, Japan\\ 
  $^{\mbox{\scriptsize\ e}}$
  Center for Computational Physics, 
  University of Tsukuba, 
  Tsukuba, Ibaraki 305-8577, Japan\\ 
  $^{\mbox{\scriptsize\ f}}$
  Department of Physics, Washington University, St. Louis, Missouri
  63130, USA\\ 
}

\date{\today}

\maketitle
\begin{abstract}
  We present a lattice NRQCD study of the $B$ meson 
  decay constant in the quenched approximation
  with emphasis given to the scaling behavior.
  The NRQCD action and the heavy-light axial current we use 
  include all terms of order $1/M$ and the perturbative
  $O(\alpha_s a)$ and $O(\alpha_s/M)$ corrections.
  Using simulations at three value of couplings 
  $\beta$=5.7, 5.9 and 6.1 on lattices of size $12^3\times 32, 
  16^3\times 48$ and $24^3\times 64$, we find
  no significant $a$ dependence in $f_B$ if the
  $O(\alpha_s a)$ correction is included in the axial current. 
  We obtain $f_B = 167(7)(15)$ MeV,
  $f_{B_s}= 191(4)(17)(^{+4}_{-0}) $ MeV and 
  $f_{B_s}/f_B =1.15(3)(1)(^{+3}_{-0})$,
  with the first error being statistical, 
  the second systematic, and the third due to 
  uncertainty of strange quark mass, while
  quenching errors being not included.
\end{abstract}

\newpage
\section{Introduction}
\label{sec:Introduction}

Lattice QCD provides a promising approach for a first-principles 
calculation of the hadronic matrix elements of $B$
meson relevant for a precision determination of 
the Cabibbo-Kobayashi-Maskawa (CKM) matrix elements.
The most important matrix elements is the $B$ meson 
leptonic decay constant $f_B$:
it is needed to determine $V_{td}$;
from the lattice technical point of view it
is the simplest $B$ meson matrix element calculable in
lattice QCD with which one can study systematic errors 
associated with a lattice treatment of heavy quark. 

The need for a careful examination of systematic errors stems from the fact 
that their magnitude for naive quark actions such as the Wilson action 
is of $O(aM)$ with $M$ the heavy quark mass. Hence errors of this origin 
can exceed 100\% for a typical lattice spacing 
of $a^{-1}\sim 2$GeV used in present simulations. 
To overcome this problem, recent lattice studies of
$f_B$~\cite{DRAPER} employ non-relativistic effective theory of 
QCD (NRQCD)~\cite{NRQCD} or a 
non-relativistic interpretation of the
relativistic lattice quark action 
for heavy quarks~\cite{EKM}.

NRQCD is an
effective theory formulated as an expansion in ${\bbox{D}}/M$ where  
$\bbox{D}$ is the spatial covariant derivative which is 
of $O(\LmQCD)$ in the heavy-light system. 
For NRQCD one has to choose the
coefficients of the expansion by imposing 
a matching condition with the full theory. This can be made 
using perturbation theory. In practice one has to truncate 
both non-relativistic expansion and perturbative expansion 
at some order so that the systematic error in NRQCD calculations 
is organized as a double expansion in $\LmQCD/M$ and the 
strong coupling constant $\alpha_s$.

An additional source of systematic errors
is the discretization error proportional to some power of $a\LmQCD$.
Since NRQCD is valid only when $aM > O(1)$,
the continuum limit $a\rightarrow 0$ can not be taken.
Therefore, removing discretization errors 
is more important in this formalism than in the usual relativistic
formulations for which continuum extrapolations can in principle be made. 
For this reason, in many lattice NRQCD calculations, the
correction terms to remove $a\LmQCD$ and even $(a\LmQCD)^2$
errors were introduced.

Until recently
the matching coefficients for the
action~\cite{NRQCDPERTDT,NRQCDPERTCM,NRQCDPERTCMKINE}
and the current operators~\cite{NRQCDPERTDTCURRENT}
have been available only at one-loop level without operator mixing.
This means that $O(\alpha_s \LmQCD/M)$ and $O(\alpha_s a\LmQCD)$ errors had
been left unremoved.
Recently, Shigemitsu and Morningstar carried out a
one-loop calculation necessary for an
$O(\alpha_s \LmQCD/M)$ and $O(\alpha_s a\LmQCD)$ improvement
of the heavy-light axial vector current~\cite{MIXING_proc,MIXING}. 
The first simulation including this improvement was
performed by Ali~Khan \textit{et al.}~\cite{SGO_paper,GLOK_paper},
in which they pointed out that the
$O(\alpha_s\LmQCD/M)$ and $O(\alpha_s a\LmQCD)$ terms significantly affect
the values of $f_B$.

The study of Ali~Khan \textit{et al.}~\cite{SGO_paper,GLOK_paper}
was made at a single lattice 
spacing corresponding to the inverse gauge coupling $\beta=6/g^2=6.0$, 
and hence left open the important question of the lattice spacing
dependence of $f_B$ obtained with lattice NRQCD (in Refs.~\cite{Hein}
Hein has calculated $f_{B_{s}}$ at $\beta$=5.7 and
discuss the scaling behavior by combining
the result at $\beta$=6.0 of Ref.~\cite{GLOK_paper}).
In this article we report on results of our systematic study 
concerning this question.
Our simulations are carried out with the plaquette action for 
gluons at $\beta=5.7, 5.9$ and 6.1 corresponding to the range of lattice 
spacing $a\sim 0.18-0.09$fm.  For light quark we employ the $O(a)$-improved 
Wilson (clover) action~\cite{CLOVER} with the tadpole improved
one-loop value for the clover coefficient~\cite{CSW,LUSCHERWEISZ}.  
We investigate in detail the effect of one-loop 
improvement of the heavy-light axial vector current as a function of 
the lattice spacing. Our main results are obtained to $O(1/M)$,
but the question of higher order corrections are examined by comparing 
the results for the NRQCD action complete to $O(1/M^2)$.

This paper is organized as follows.
In Sec.~\ref{sec:Lattice_NRQCD_Action} we summarize the NRQCD action 
we use. In Sec.~\ref{sec:Improvement_of_the_Current} 
improvement of the axial vector current is discussed, 
and our one-loop mixing coefficients are presented. 
Details of the simulations and our methods for extraction of the 
decay constant are given in Sec.~\ref{sec:Details_of_the_Simulation}
together with numerical results.  
We discuss the effect of improvement in the static limit 
in Sec.~\ref{sec:Static_Limit}.  
Our results for $f_B$ are presented in Sec.~\ref{sec:B_Meson_Decay_Constant}
where a comparison is also made with those obtained 
with the relativistic formalism.
In Sec.~\ref{sec:Mass_Splittings} the hyperfine splitting of 
the $B$ meson and the $B_s-B$ mass difference are given. 
Our conclusions are summarized in Sec.~\ref{sec:Conclusions}.

\section{Lattice NRQCD Action}
\label{sec:Lattice_NRQCD_Action}
\subsection{Form of action}

Let us denote by  $Q(t,\bbox{x})$ the two-component heavy quark field.
This field evolves in the time direction according to the action,
\begin{equation}
  \label{eqn:A}
  S = \sum_{t,\bbox{x}} Q^{\dag}(t,\bbox{x})
  \left[ Q(t,\bbox{x}) - K_{t}Q(t-1,\bbox{x}) \right].
\end{equation}
where the operator $K_t$ specifies the evolution; our choice is
\begin{equation}
  K_{t} = 
  \left( 1-\frac{a H_{0}}{2 n} \right)^{n}_{t}
  \left( 1-\frac{a \delta H}{2} \right)_{t} 
  {U^{\dagger}_{4}}_{t-1}
  \left( 1-\frac{a \delta H}{2} \right)_{t-1}
  \left( 1-\frac{a H_{0}}{2 n} \right)^{n}_{t-1},
  \label{eqn:ACT}
\end{equation}
Here subscripts represent the time slice at which Hamiltonian 
operators such as $(1- a H_{0}/2n)$ act, and an integer 
$n$ is introduced to avoid instability appearing in the 
evolution equation due to unphysical momentum modes~\cite{NRQCD}.
We note that the ordering of terms in Eq.~(\ref{eqn:ACT}) is different
from the one employed in~\cite{GLOK_paper}: 
the factor $(1-a \delta H/2)$ is placed 
inside of $(1-a H_{0}/2n)$ in our choice.

The leading order Hamiltonian $H_{0}$ is given by 
\begin{eqnarray}
  H_{0} & = & -\frac{\Dtwo}{2 M_{0}}. 
\end{eqnarray}
For the correction term $\delta H$, we consider two choices 
corresponding to the non-relativistic expansion to order $1/M$
($\delta H_{\mbox{\scriptsize I}}$) or  
to order $1/M^2$ ($\delta H_{\mbox{\scriptsize II}}$), given by 
\begin{eqnarray}
  \delta H_{\mbox{\scriptsize I}} 
  & = & - c_{1}\frac{g}{2 M_{0}}\bbox{\sigma}\cdot\bbox{B}, \\
  \delta H_{\mbox{\scriptsize II}}
  & = & - c_{1}\frac{g}{2 M_{0}}\bbox{\sigma}\cdot\bbox{B}
  + c_{2}\frac{i g}{8M^{2}_{0}}(\Done\cdot\bbox{E}-\bbox{E}\cdot\Done)
  \nonumber \\
  & &
  - c_{3}\frac{g}{8M^{2}_{0}}\bbox{\sigma}\cdot
         (\Done\times\bbox{E}-\bbox{E}\times\Done)
  \nonumber \\
  & &
  - c_{4}\frac{(\Dtwo)^{2}}{8M^{3}_{0}}
  + c_{5}\frac{a^{2} \bbox{\Delta}^{(4)}}{24M_{0}}
  - c_{6}\frac{a(\Dtwo)^{2}}{16nM^{2}_{0}}.
\label{eq:correction}
\end{eqnarray}
We refer to the two choices as NRQCD-I and NRQCD-II.
We work with both Hamiltonians in parallel and compare their
results in order to examine effects of truncation in the $1/M$ expansion. 
Various covariant differential operators in the Hamiltonian
are defined in terms of the forward and backward derivatives 
$\Delta^{(+)}_{\mu}$ and $\Delta^{(-)}_{\mu}$ in the $\mu$-th direction
as
$\Delta^{(\pm)}_{\mu} \equiv (\Delta^{(+)}_{\mu}+\Delta^{(-)}_{\mu})/2$,
$\Delta^{(2)}_{\mu} \equiv 
 \Delta^{(+)}_{\mu} \Delta^{(-)}_{\mu}$,
$\Dtwo \equiv \sum_{i=1}^{3} \Delta^{(2)}_{i}$, and
$\bbox{\Delta}^{(4)} \equiv \sum_{i=1}^{3}(\Delta^{(2)}_{i})^{2}$. 
The field strength operators $\bbox{B}$ and $\bbox{E}$ are
constructed with the clover-leaf definition as in Ref.~\cite{NRQCD}.  
The bare heavy quark mass is denoted as $M_{0}$, and $c_i$'s
are parameters to describe the strength of each term.

The relativistic four-component field $\psi_{h}$ is related
to the effective field $Q$ through the Foldy-Wouthuysen-Tani (FWT) 
transformation, 
\begin{equation}
  \label{eqn:FWT}
  \psi_{h}(t,\bbox{x}) = R Q(t,\bbox{x}),
\end{equation}
Here the transformation operator $R$ is given by
\begin{eqnarray}
  R_{\mbox{\scriptsize I}}  & = &
  1 - d_{1} \frac{\bbox{\gamma}\cdot\Done}{2M_{0}}, \\
  R_{\mbox{\scriptsize II}} & = &
  1 - d_{1} \frac{\bbox{\gamma}\cdot\Done}{2M_{0}}
    + d_{2} \frac{\Dtwo}{8 M^{2}_{0}}  
  \nonumber \\
  & & \ \ \ 
    + d_{3} \frac{g}{8M^{2}_{0}}\bbox{\Sigma}\cdot\bbox{B}  
    - d_{4} \frac{i g}{4M^{2}_{0}}\gamma_{4}\bbox{\gamma}\cdot\bbox{E},
\end{eqnarray}
with $\Sigma^{j}= \mbox{diag}\{\sigma^{j},\sigma^{j}\}$, 
and $R_{\mbox{\scriptsize I}}$ ($R_{\mbox{\scriptsize II}}$) is 
to be used in conjunction with $\delta H_{\mbox{\scriptsize I}}$ 
($\delta H_{\mbox{\scriptsize II}}$) to achieve the desired
accuracy in the $1/M$ expansion. 

The coefficients $c_{i}$ and $d_{i}$ should be determined
by matching the action to the continuum relativistic QCD
action by either resorting to perturbation theory 
or estimating it non-perturbatively so as to
reproduce the same theory in each order of the $1/M$ expansion. 
So far even perturbative results are not available for these coefficients. 
We adopt the tree-level value $c_{i}=1$ and $d_{i}=1$ in our work, 
applying, however, the mean-field improvement to all the link 
variables in the action and the FWT transformation with the replacement
$U_{\mu}\rightarrow U_{\mu}/u_{0}$, where we take
$u_{0}=\langle 
       \mbox{Tr} U_{\mbox{\scriptsize plaq}}/3
       \rangle^{1/4}$~\cite{LEPAGEMACKENZIE}.

\section{Improvement of the Current}
\label{sec:Improvement_of_the_Current}

To calculate the decay constant $f_B$, 
the heavy-light axial vector current in lattice NRQCD has to be matched 
to that in continuum QCD.
For the overall renormalization factor $Z_A$ 
this was first performed by Davies and Thacker~\cite{NRQCDPERTDTCURRENT}
by perturbation theory to one-loop order.
An important recent development made by Shigemitsu and 
Morningstar~\cite{MIXING_proc,MIXING} is that the matching has been extended to 
$O(\alpha_s a\LmQCD)$ and $O(\alpha_s \LmQCD/M)$. 
Since our choice of the action is slightly different from that 
used by the authors of Refs.~\cite{NRQCDPERTDTCURRENT,MIXING}, 
we have repeated the one-loop calculation 
for our NRQCD action.

Consider the axial vector current  
${A_4}_{\mbox{\scriptsize cont}}$ in the continuum. 
We demand that on-shell $S$ matrix elements of the lattice 
axial current reproduce that of the continuum current 
up to $O(\bbox{p})$ with $\bbox{p}$ the spatial
momentum of the heavy or light quark. 
At one-loop level the relation takes the form
\begin{eqnarray}
  {A_{4}}_{\mbox{\scriptsize cont}} & = & 
  \left[ 1 + \alpha_{s} \rho^{(0)}_{A} \right] 
  J^{(0)}_{\mbox{\scriptsize latt}}
  + \alpha_{s}\rho^{(1)}_{A} J^{(1)}_{\mbox{\scriptsize latt}} 
  + \alpha_{s}\rho^{(2)}_{A} J^{(2)}_{\mbox{\scriptsize latt}},
  \label{eq:FULLA4}
\end{eqnarray}
where the heavy-light lattice operators of dimension 3
and 4 are defined by 
\begin{eqnarray}
  J^{(0)}_{\mbox{\scriptsize latt}} & = & 
    \bar{\psi}_{l} \Gamma \psi_{h},
  \label{eqn:J0} \\
  J^{(1)}_{\mbox{\scriptsize latt}} & = &
    \frac{-1}{2M_{0}}\bar{\psi}_{l} \Gamma
    \bbox{\gamma}\cdot \Done \psi_{h},
  \label{eqn:J1} \\
  J^{(2)}_{\mbox{\scriptsize latt}} & = &
    \frac{1}{2M_{0}}
    \bar{\psi}_{l}\bbox{\gamma}\cdot\loarrow{\bbox{\Delta}}^{(\pm)}
    \Gamma \psi_{h}
  \label{eqn:J2},
\end{eqnarray}
with $\Gamma=\gamma_{5}\gamma_{4}$ for the temporal
axial vector current, and 
$\psi_l$ and $\psi_h$ denoting the light and heavy quark
fields, respectively.
We calculate the  coefficients $\rho_A^{(i)}$ 
for NRQCD-I for heavy quark and 
the $O(a)$-improved clover action~\cite{CLOVER}
for the light quark.  The use of clover action for
the light quark is necessary to achieve the accuracy of
$O(\alpha_s a)$ in matching the current.
For renormalization of the continuum current we adopt 
the $\overline{\mbox{MS}}$ scheme 
using dimensional regularization with fully anti-commuting 
$\gamma_5$. We also apply the tadpole improvement
procedure~\cite{LEPAGEMACKENZIE} with
the average plaquette for all link variables in the
covariant derivative of the operators, (\ref{eqn:J1}) and (\ref{eqn:J2}),
and with the critical hopping
parameter for the wave function renormalization of the light quark
fields consistently in both non-perturbative and perturbative
calculations.

Numerical results for the coefficients $\rho_A^{(i)}$ are listed in
Table~\ref{tab:one-loop_coefficients}, and plotted in 
Figure~\ref{fig:one-loop_coefficients} as a function of $1/aM_{0}$.
For $\rho_A^{(0)}$ the difference $\rho_A^{(0)}-(1/\pi)\ln(aM_0)$ is 
shown since the coefficient of the leading operator 
$1+\alpha_s\rho_A^{(0)}$, being the usual renormalization factor $Z_A$, 
contains a logarithmic term $(1/\pi)\ln(aM_0)$.
The other coefficients $\rho_A^{(1)}$ and $\rho_A^{(2)}$ are 
divided by $2aM_0$.
The filled symbols represent the values explicitly obtained 
with the static action~\cite{STATICMIXING}.
We have confirmed that the infinite mass limit of
$\rho_A^{(0)}-(1/\pi)\ln(aM_0)$ agrees with the static results
of Borrelli and Pittori~\cite{Borrelli_Pittori_92} and of
Golden and Hill~\cite{Golden_Hill_91}.

We observe that $\rho_A^{(1)}/2aM_0$ vanishes in the
limit $aM\rightarrow\infty$, which tells us that the contribution
of $\alpha_s\rho^{(1)}_{A}J^{(1)}_{\mbox{\scriptsize latt}}$ is of 
$O(\alpha_s \LmQCD/M)$.
This is expected since $J^{(1)}_{\mbox{\scriptsize latt}}$
involves a derivative of the heavy quark field.
On the other hand, $\alpha_s\rho_A^{(2)}J^{(2)}_{\mbox{\scriptsize latt}}$
does not contain such a derivative, and 
$\rho_A^{(2)}/2aM_0$ remains finite in the 
static limit as seen in Figure~\ref{fig:one-loop_coefficients}.
Thus its contribution contains terms of $O(\alpha_s a\LmQCD)$.
This term is an analogue of the current improvement term of
$O(\alpha_s a)$ for the light quark discussed in Ref.~\cite{LUSCHERWEISZ}. 

We add a remark that we have repeated the one-loop calculation
for the action employed in Ref.~\cite{MIXING}, and numerically confirmed
their results to a 3 digit accuracy.

\section{Details of the Simulation}
\label{sec:Details_of_the_Simulation}

\subsection{Run Parameters}
\label{subsec:Actions_and_Parameters}

We list our simulation parameters in Table~\ref{tab:lattice_parameters}.
Our simulations are carried out for three values of the 
coupling $\beta$=5.7, 5.9 and 6.1 using the standard plaquette action
for gluons.
The corresponding values of the lattice spacing $a$ is about 
0.18, 0.13 and 0.09 fm, respectively,
if determined from the string tension.
We choose our spatial lattice size to be larger than 2 fm.

For heavy quark we take five values of the bare mass $aM_0$ for 
each $\beta$ to cover a range of the physical heavy quark mass $M$ 
between 2 GeV and 16 GeV.
This wide range of heavy quark mass enables us to examine explicitly
the $1/M$ dependence of $f_B$.
The parameter $n$ is chosen so as to satisfy the stability condition 
$n > 3/aM_{0}$. 

For light quark we use the $O(a)$-improved Wilson
action~\cite{CLOVER} with the clover coefficient
$c_{\mbox{\scriptsize sw}}=(1/u_0^3)[1+0.199\alpha_{V}(1/a)]$,
which includes the $O(\alpha_s)$ correction calculated 
in Refs.~\cite{CSW,LUSCHERWEISZ}.
Four values of the light quark hopping parameter $\kappa$ 
are employed for extrapolation to the chiral limit (see 
Table~\ref{tab:lattice_parameters} for numerical values).

In the quenched approximation the value of the strange quark mass $m_s$
differs depending on whether $m_K$ or $m_{\phi}$
is used as input. 
The value of $m_s$ determined with $m_{\phi}$ is 
higher than that with $m_K$, and the discrepancy does 
not become smaller for smaller lattice spacings.
We choose to calculate $f_{B_s}$ for both $m_s$, 
and take their difference as a systematic error.
The hopping parameters $\kappa_s$ 
($\kappa_{s1}$ from $m_K$ and $\kappa_{s2}$ from $m_{\phi}$)
are also given in 
Table~\ref{tab:lattice_parameters}. 

The physical scale of lattice spacing $a$ is fixed using the string 
tension $\sigma=427$ MeV. 
Recent data of the string tension for the standard plaquette
action are summarized in Ref.~\cite{string_tension}.
We adopt their parameterization to 
obtain the values of $1/a$ at our $\beta$.

\subsection{Fitting procedure and data analysis}

The method to extract the heavy-light decay constant is
standard.
We define a local and a smeared operator for the
pseudoscalar channel by
\begin{eqnarray}
  O^{\mbox{\scriptsize L}}_{P}(t,\bbox{x}) & = &
  \bar{\psi}_{l}(t,\bbox{x})\gamma_{5}\psi_{h}(t,\bbox{x}),
  \label{eqn:PL}\\
  O^{\mbox{\scriptsize S}}_{P}(t,\bbox{x}) & = &
  \sum_{\bbox{y}}
  \bar{\psi}_{l}(t,\bbox{x})\gamma_{5}\psi_{h}(t,\bbox{y})
  \phi^{\mbox{\scriptsize SRC}}(|\bbox{x}-\bbox{y}|),
  \label{eqn:PS}
\end{eqnarray}
in the Coulomb gauge.
For the smearing function we use the form
$\phi^{\mbox{\scriptsize SRC}}(|\bbox{x}|)
 = \exp( - a |\bbox{x}|^{b} )$,
with the parameters $a$ and $b$ chosen so as to reproduce
the functional form of the heavy-light meson wave function
measured in our simulations.
We measure the two-point functions given by 
\begin{eqnarray}
  C^{\mbox{\scriptsize LS}}_{PP}(t_{f},t_{i}) & = &
  \sum_{\bbox{x}_f}
  \langle O^{\mbox{\scriptsize L}}_{P}(t_{f},\bbox{x}_{f})
         {O^{\mbox{\scriptsize S}}_{P}}^{\dag}(t_{i},\bbox{0})\rangle,
  \label{eqn:CLSPP}\\
  C^{\mbox{\scriptsize SS}}_{PP}(t_{f},t_{i}) & = &
  \sum_{\bbox{x}_f}
  \langle O^{\mbox{\scriptsize S}}_{P}(t_{f},\bbox{x}_{f})
         {O^{\mbox{\scriptsize S}}_{P}}^{\dag}(t_{i},\bbox{0})\rangle,
  \label{eqn:CSSPP}\\
  C^{\mbox{\scriptsize LS}}_{J^{(i)}P}(t_{f},t_{i}) & = &
  \sum_{\bbox{x}_f}
  \langle J^{(i)}_{\mbox{\scriptsize latt}}(t_{f},\bbox{x}_{f})
  {O^{\mbox{\scriptsize S}}_{P}}^{\dag}(t_{i},\bbox{0})\rangle,
  \label{eqn:CLSJP}
\end{eqnarray}
with the Dirichlet boundary condition in temporal direction.
In this measurement the source is placed at the time slice $t_i$=6
(at $\beta$=5.7), 7 (5.9) and 16 (6.1).
For the heavy-light meson with zero spatial momentum, 
$C^{\mbox{\scriptsize LS}}_{J^{(1)}P}(t_f,t_i)$ and
$C^{\mbox{\scriptsize LS}}_{J^{(2)}P}(t_f,t_i)$ are
identical by construction. 

We fit the correlators to the exponential form, 
\begin{eqnarray}
  C^{\mbox{\scriptsize LS}}_{PP}(t_f,t_i)
  & \rightarrow &
  Z^{\mbox{\scriptsize LS}}_{PP} 
  \exp( -a E^{\mbox{\scriptsize bin}} (t_f-t_i) ),
  \label{eqn:CLSPPFIT}\\
  C^{\mbox{\scriptsize SS}}_{PP}(t_f,t_i)
  & \rightarrow &
  Z^{\mbox{\scriptsize SS}}_{PP} 
  \exp( -a E^{\mbox{\scriptsize bin}} (t_f-t_i) ),
  \label{eqn:CSSPPFIT}\\
  C^{\mbox{\scriptsize LS}}_{J^{(i)}P}(t_f,t_i)
  & \rightarrow &
  Z^{\mbox{\scriptsize LS}}_{J^{(i)}P} 
  \exp( -a E^{\mbox{\scriptsize bin}} (t_f-t_i) ),
  \label{eqn:CLSJPFIT}
\end{eqnarray}
over a range of $t$ where we find a plateau in the
effective mass plot. 
Representative effective mass plots for 
$C^{\mbox{\scriptsize LS}}_{PP}(t_f,t_i)$, 
$C^{\mbox{\scriptsize SS}}_{PP}(t_f,t_i)$,
$C^{\mbox{\scriptsize LS}}_{J^{(0)}P}(t_f,t_i)$, and 
$C^{\mbox{\scriptsize LS}}_{J^{(1)}P}(t_f,t_i)$ are shown in 
Figures~\ref{fig:effective_mass_heaviest} 
(\ref{fig:effective_mass_lightest})
for the case of the heaviest (lightest) quark masses at $\beta$=6.1.
The signal is remarkably clean even for 
$C^{\mbox{\scriptsize LS}}_{J^{(1)}P}$ which
includes a spatial differential operator.
To constrain the fit as much as possible we take the binding energy 
$E^{\mbox{\scriptsize bin}}$ to be common among the correlators.
This is particularly necessary for a stable extraction of
$Z^{\mbox{\scriptsize SS}}_{PP}$ since the signal for
$C^{\mbox{\scriptsize SS}}_{PP}(t_f,t_i)$ is much noisier
than those for the others. 
We estimate statistical errors of the fitted parameters 
using the jackknife method with unit bin size. 
Statistical correlation of data between different time
slices or between different mass parameters is neglected in
the fitting.

\subsection{Heavy-light Meson Mass}
\label{sec:Self_Energy_Corrections}

We calculate the pseudoscalar meson mass $aM_{P}$ from a sum of the
renormalized heavy quark mass and the binding energy
$E^{\mbox{\scriptsize bin}}$ through the formula
\begin{equation}
  aM_{P} = Z_{m} aM_{0} - E + a E^{\mbox{\scriptsize bin}},
\label{eqn:MESONMASS}
\end{equation}
where $E$ is the energy shift and $Z_{m}$ the kinetic mass 
renormalization of the heavy quark.

The one-loop calculation of $E$ and $Z_m$ was first carried out 
by Davies and Thacker~\cite{NRQCDPERTDT} and by 
Morningstar~\cite{NRQCDPERTCM}.
We repeat the calculation for our action NRQCD-I.
We write the perturbative expansion of $E$, $Z_m$ and the wave function
renormalization $Z_{2h}$ as
\begin{eqnarray}
  E      & = &      \alpha_{s} A, \\
  Z_{m}  & = &  1 + \alpha_{s} B, \\
  Z_{2h} & = &  1 + \alpha_{s} C,
\end{eqnarray}
and list $A$, $B$, and $C$ in 
Table~\ref{tab:one-loop_coefficients}.

\subsection{Heavy-light Decay Constant}
\label{sec:Meson_Mass_and_Decay_Constant}

The pseudoscalar meson decay constant $f_P$ is constructed from
the contribution of each operator
$J^{(i)}_{\mbox{\scriptsize latt}}$ defined by 
\begin{equation}
  a^{3/2} (f_P\sqrt{M_P})^{(i)} =
  \frac{a^{3/2}}{\sqrt{M_P}} 
  \langle 0|J^{(i)}_{\mbox{\scriptsize latt}}|P\rangle = 
  Z^{\mbox{\scriptsize LS}}_{J^{(i)}P}
  \sqrt{\frac{2}{Z^{\mbox{\scriptsize SS}}_{PP}}}
  \sqrt{1-\frac{3\kappa}{4\kappa_{\mbox{\scriptsize crit}}}},
\end{equation}
where $\sqrt{1-3\kappa/4\kappa_{\mbox{\scriptsize crit}}}$
represents the tadpole-improved wave function normalization
factor for light quark. 
Including the one-loop corrections, the decay constant is
given by
\begin{equation}
  a^{3/2} (f_P\sqrt{M_P}) = 
  [1+\alpha_s\rho_A^{(0)}] a^{3/2} (f_P\sqrt{M_P})^{(0)}
  + \sum_{i=1}^{2} 
     \alpha_s\rho_A^{(i)}  a^{3/2} (f_P\sqrt{M_P})^{(i)}.
  \label{eq:fsqrtm}
\end{equation}
We note that 
$a^{3/2} (f_P\sqrt{M_P})^{(1)}=a^{3/2}(f_P\sqrt{M_P})^{(2)}$
holds in the rest frame of the heavy-light meson.

In Figures~\ref{fig:chiral_limit_binding_energy} and
\ref{fig:chiral_limit_decay_constant} we show 
$aE^{\mbox{\scriptsize bin}}$ and $a^{3/2}(f_P\sqrt{M_P})^{(i)}$ 
as a function of $1/\kappa$ together with a linear (solid lines) 
and a quadratic (dotted lines) fit.  We employ the linear fit 
for chiral extrapolation since the difference between the linear and 
quadratic fits are negligibly small compared with 
the errors of the data.
The linear fit is also used for an interpolation to the strange quark. 
The values of $aE^{\mbox{\scriptsize bin}}$ and
$a^{3/2}(f_P\sqrt{M_P})^{(i)}$ at 
$\kappa=\kappa_{\mbox{\scriptsize crit}}$ as
well as those at $\kappa_{s1}$ and $\kappa_{s2}$ extracted in this 
way are summarized in 
Tables~\ref{tab:ps_mass}, \ref{tab:fsqrtm_rawdata_f^0}
and \ref{tab:fsqrtm_rawdata_f^1}. 

One of the points we discuss in detail below is the effect of 
$O(\alpha_s a\LmQCD)$ improvement 
in the static limit. For this purpose 
we need to extract the decay constant in the static limit.

Figure~\ref{fig:fsqrtm_vs_1/M_P} shows the dependence of
$(f_P\sqrt{M_P})^{(0)}$ as a function of $1/M_P$ for each
$\beta$ where $M_P$ is calculated by the tree-level formula. 
The physical scale of lattice spacing $a$ is determined 
from the string tension. We fit the mass dependence by 
\begin{equation}
  \label{eq:1/M_P_dependence_of_fsqrtm0}
   a^{3/2}(f_P\sqrt{M_P})^{(0)} =
  \left.  a^{3/2}  (f_P\sqrt{M_P})^{(0)} \right|_{\mbox{\scriptsize static}}
  \left( 1 + \frac{a_1}{aM_P} + \frac{a_2}{(aM_P)^2} \right).
\end{equation}
We also fit the mass dependence 
of $2aM_{0} a^{3/2}(f_P\sqrt{M_P})^{(1)}$ by 
\begin{equation}
  \label{eq:1/M_P_dependence_of_fsqrtm1}
         2aM_{0} a^{3/2}(f_P\sqrt{M_P})^{(1)} =
  \left. 2aM_{0} a^{3/2}(f_P\sqrt{M_P})^{(1)} \right|_{\mbox{\scriptsize static}}
  \left( 1 + \frac{a'_1}{aM_P} + \frac{a'_2}{(aM_P)^2} \right).
\end{equation}
Similarly, Figure~\ref{fig:fsqrtm^1/fsqrtm^0} shows the
$1/M_P$ dependence of the ratio 
$-2M_0 (f_P\sqrt{M_P})^{(1)}/(f_P\sqrt{M_P})^{(0)}$.
The functional dependence can also be parameterized as
\begin{equation}
  \label{eq:1/M_P_dependence_of_fsqrtm1/fsqrtm0}
  -2aM_0 \frac{(f_P\sqrt{M_P})^{(1)}}{(f_P\sqrt{M_P})^{(0)}}
  =
  -2aM_0 \left. 
    \frac{(f_P\sqrt{M_P})^{(1)}}{(f_P\sqrt{M_P})^{(0)}}
    \right|_{\mbox{\scriptsize static}} 
  \left(1 + \frac{b_1}{aM_P} + \frac{b_2}{(aM_P)^2} \right).
\end{equation}
The values of
$a^{3/2}(f_P\sqrt{M_P})^{(0)}|_{\mbox{\scriptsize static}}$ and 
$2aM_{0} a^{3/2}(f_P\sqrt{M_P})^{(1)}|_{\mbox{\scriptsize static}}$  
are given in Table~\ref{tab:fsqrtm_rawdata_f^0} and
\ref{tab:fsqrtm_rawdata_f^1}, respectively.

To obtain the $B$ meson decay constant at the physical $B$
meson mass, we fit the $1/M_P$ dependence of 
the renormalized quantity $f_P\sqrt{M_P}$ in the
renormalization group invariant form
$\Phi_P \equiv [\alpha_s(M_P)/\alpha_s(M_B)]^{2/11} 
               f_P\sqrt{M_P}$
instead of fitting the contribution of individual operators 
and summing the results.

\section{Static Limit}
\label{sec:Static_Limit}

We begin discussion of our results with the lattice spacing dependence
of the decay constant in the static limit.
This limit has the advantage that errors that depend on
the heavy quark mass 
such as $O(\alpha_s\LmQCD/M)$ vanish, and hence we can see
the effect of $O(\alpha_s a\LmQCD)$ improvement more clearly.

According to the discussion in Sec.~\ref{sec:Improvement_of_the_Current},
the contribution of
$J^{(1)}_{\mbox{\scriptsize latt}}$ vanishes in the static limit. 
From Eq.~(\ref{eq:FULLA4}), the matching relation in the static limit
for the axial vector current is given by
\begin{eqnarray}
  {A_{4}}_{\mbox{\scriptsize cont}} & = & 
  \left[ 1 + \alpha_{s}\rho^{(0)}_{\mbox{\scriptsize static}} \right] 
  J^{(0)}_{\mbox{\scriptsize static}}
  + \alpha_{s}\rho^{(\mbox{\scriptsize disc})}_{\mbox{\scriptsize static}} 
  a J^{(\mbox{\scriptsize disc})}_{\mbox{\scriptsize static}},
  \label{eq:A4_static}
\end{eqnarray}
where $\rho^{(0)}_{\mbox{\scriptsize static}}$ and  $J^{(0)}_{\mbox{\scriptsize  static}}$ are
the naive static limit (except anomalous dimension) of 
$\rho^{(0)}_{A}$ and $J^{(0)}_{\mbox{\scriptsize latt}}$.
$\rho^{(\mbox{\scriptsize disc})}_{\mbox{\scriptsize static}}$ and $J^{(\mbox{\scriptsize disc})}_{\mbox{\scriptsize static}}$ are defined as
\begin{eqnarray}
\rho^{(\mbox{\scriptsize disc})}_{\mbox{\scriptsize static}}&=&\lim_{aM_{0}\rightarrow \infty}
 \rho_A^{(2)}/2aM_0, \\
  aJ^{(\mbox{\scriptsize disc})}_{\mbox{\scriptsize static}}&=&\lim_{aM_{0}\rightarrow \infty}
2aM_0 J^{(2)}_{\mbox{\scriptsize latt}}.
\end{eqnarray}
The numerical value of the matching coefficients 
in the static limit is given in 
Table~\ref{tab:one-loop_coefficients}.

The decay constant is calculated from
\begin{equation}
  f_{B_{(s)}}^{\mbox{\scriptsize  static}} \equiv
 (f_{P_{(s)}}\sqrt{M_{P_{(s)}}})|_{\mbox{\scriptsize  static}}/\sqrt{M_{B_{(s)}}}
\end{equation}
with 
\begin{eqnarray}
  \label{eq:fsqrtm_static}
  \lefteqn{ \left.(f_P\sqrt{M_P})\right|_{\mbox{\scriptsize  static}} }
  \nonumber \\ 
  & = & 
  \left[ 1 + \alpha_{s} \rho^{(0)}_{\mbox{\scriptsize  static}} \right] 
  \left\{\left.(f_P\sqrt{M_P})^{(0)}\right|_{\mbox{\scriptsize  static}}\right\}
  + \alpha_{s}\rho^{(\mbox{\scriptsize disc})}_{\mbox{\scriptsize  static}} 
  \left\{\left.2aM_0 (f_P\sqrt{M_P})^{(1)}\right|_{\mbox{\scriptsize  static}}\right\}.
\end{eqnarray}
A nominal value of $M_0$=4.5 GeV is used for the heavy quark mass 
to evaluate the logarithm of $\rho_{\mbox{\scriptsize  static}}^{(0)}$.
For the strong coupling constant 
$\alpha_s$ we employ $\alpha_V(q^{\ast})$~\cite{LEPAGEMACKENZIE}
evolved from $\mu=3.40/a$ to $q^{\ast}$.
Within one-loop calculations
there is an uncertainty in the choice of the scale
$q^{\ast}$.
We take the average of results obtained with
$q^{\ast}=\pi/a$ and with $1/a$, and consider the discrepancy from
the two choices of $q^{\ast}$ as an upper and lower bound for the error 
due to two-loop corrections in the renormalization factor.

Figure~\ref{fig:fsqrtm_static} shows the $a$ dependence of
the decay constant in the static limit $f_B^{\mbox{\scriptsize  static}}$ and 
$f_{B_{s}}^{\mbox{\scriptsize  static}}$. 
Open symbols represent the results which are not corrected for
the mixing  effect of the operator
$a J^{(\mbox{\scriptsize disc})}_{\mbox{\scriptsize  static}}$
( which corresponds to the static limit of
$2aM_0J^{(2)}_{\mbox{\scriptsize latt}}$),
 and filled symbols include
this effect.  Statistical errors are shown with solid bars, and 
uncertainties due to the choice of $q^{\ast}$ by dotted bars.
From the figure
we see that an apparent $a$ dependence for the unimproved 
results is removed by the inclusion of the higher dimensional operator
$J^{(\mbox{\scriptsize disc})}_{\mbox{\scriptsize  static}}$
at the one-loop level.

A worry with this observation is a sizable systematic error
due to two-loop uncertainties. 
On this point we note that an estimate for the optimal value of $q^{\ast}$ 
for the multiplicative renormalization coefficient is known to be 
$q^{\ast}=2.18/a$ for the combination of
the static heavy quark and the unimproved Wilson light 
quark~\cite{Hernandez-Hill}.  Since there seems to be no obvious 
reason that this value changes significantly for the $O(a)$-improved 
light quark action, taking the difference of the results for $q^{\ast}=\pi/a$ 
and $1/a$ may well be an overestimate of the two-loop uncertainty. 
An alternative estimate employing $\alpha_s(2/a)$ would reduce the 
error estimate by roughly a factor two.  
Furthermore the magnitude of this error is correlated among different 
$\beta$, and between results without and with the improvement
at each $\beta$.   Hence a reduction of 
the $a$ dependence is less affected by the choice
of $q^{\ast}$ than $f_{B_{(s)}}^{\mbox{\scriptsize  static}}$
itself.

Numerically the magnitude of the $O(\alpha_s a)$ term relative to 
that of the leading operator $J^{(0)}_{\mbox{\scriptsize static}}$ 
(the static limit of $J^{(0)}_{\mbox{\scriptsize latt}}$)
takes values, 
\begin{eqnarray}
  \lefteqn{
    \alpha_{s}
    \times 
    \rho^{(\mbox{\scriptsize disc})}_{\mbox{\scriptsize  static}}
    \times 
    \left.
    2aM_0\frac{(f_P\sqrt{M_P})^{(1)}}{(f_P\sqrt{M_P})^{(0)}}
    \right|_{\mbox{\scriptsize  static}}
    }
  \nonumber \\
  & = & 0.272(84) \times 1.036 \times (-0.712(9)) = -0.201(84)
  \;\;\;\mbox{at $\beta$=5.7}, \nonumber \\ 
  & = & 0.217(53) \times 1.036 \times (-0.622(7)) = -0.140(53)
  \;\;\;\mbox{at $\beta$=5.9}, \nonumber \\
  & = & 0.189(40) \times 1.036 \times (-0.546(8)) = -0.107(41)
  \;\;\;\mbox{at $\beta$=6.1},
\end{eqnarray}
where the error is dominated by the uncertainty in
$\alpha_s$. 
At $\beta=6.1$ the effect reduces $f_{B}^{\mbox{\scriptsize  static}}$
by about 10\% from the value without the improvement term.

\section{$B$ Meson Decay Constant}
\label{sec:B_Meson_Decay_Constant}

\subsection{Dependence on heavy-light meson mass}
\label{sec:1/M_P_dependence}

In Figure~\ref{fig:Phi_vs_1/M_P} we present  
$\Phi_P \equiv [\alpha_s(M_P)/\alpha_s(M_B)]^{2/11} 
               f_P\sqrt{M_P}$
as a function of $1/M_P$ for three values of $\beta$ of our simulation. 
Open symbols denote results from the leading operator alone, and 
filled symbols show how they change due to the inclusion
of the higher dimensional operators 
$J^{(1)}_{\mbox{\scriptsize latt}}$ and 
$J^{(2)}_{\mbox{\scriptsize latt}}$.
The factor $[\alpha_s(M_P)/\alpha_s(M_B)]^{2/11}$ is
introduced to cancel the logarithmic divergence
$(1/\pi)\ln(aM_0)$ in the one-loop coefficient
$\rho_A^{(0)}$. 
For $\alpha_s(M_P)$ we use $\alpha_V(\mu)$~\cite{LEPAGEMACKENZIE}
evolved from $\mu=3.40/a$ to $M_P$. 
The chiral limit is taken for the light quark.
Solid and dotted error bars show the statistical error
and the uncertainty due to two-loop corrections in the renormalization 
factors.  The latter is estimated in the same way 
as for the static limit discussed in Sec.~\ref{sec:Static_Limit}.

As first observed in Refs.~\cite{MIXING_proc,SGO_paper,GLOK_paper}, 
the contributions from the operators 
$J^{(1)}_{\mbox{\scriptsize latt}}$ and
$J^{(2)}_{\mbox{\scriptsize latt}}$ sizably affects
the decay constant.   The dominant effect arises from 
$J^{(2)}_{\mbox{\scriptsize latt}}$. 
A larger difference between the two sets of results toward the 
static limit is explained by the fact that
the one-loop coefficient $\rho_A^{(2)}/2aM_0$ increases 
larger toward this limit (see Figure~\ref{fig:one-loop_coefficients}). 
In contrast, the contribution of $J^{(1)}_{\mbox{\scriptsize latt}}$ is 
negligible since the
perturbative coefficient $\rho_A^{(1)}/2aM_0$ stays
very small ($|\rho_A^{(1)}/2aM_0|<$0.2) for our
heavy quark mass $aM_0>$1.2.

As was the case for the decay constant in the static limit, 
uncertainties due to two-loop corrections are sizable, particularly 
at $\beta$=5.7. 
This uncertainty does decrease, however, for weaker couplings 
at $\beta=5.9$ and 6.1. It also becomes smaller 
as one moves down from the static limit toward the physical point 
for the $B$ meson $M_P=M_B$. 

\subsection{Dependence on lattice spacing}
\label{sec:a_dependence}

By interpolating data shown in Figure~\ref{fig:Phi_vs_1/M_P} to 
the physical $B$ meson mass, we obtain $f_B$ for each $\beta$.
The decay constant $f_{B_s}$ for $B_s$ meson is calculated in a similar
manner. 
The bare $b$ quark mass that reproduces the physical $B$
meson mass is listed in Table~\ref{tab:b-quark_mass}, and
$f_B$ and $f_{B_s}$ at each $\beta$ are 
given in Table~\ref{tab:decay_constant_results} for the 
two choices of the scale $q^{\ast}=\pi/a$ and $1/a$. 

The lattice spacing dependence of $f_B$ and $f_{B_s}$ is
shown in Figure~\ref{fig:f_B_vs_a}.
Looking at the central values, we observe that a large $a$ dependence 
exhibited by the results without the operator mixing (open symbols) is
removed in the full result (filled symbols).  This feature is clearer
for $f_{B_s}$; a variation is seen for $f_B$ between $\beta=5.9$ and 6.1, 
albeit with simultaneously larger statistical errors. 
While we need to keep in mind the uncertainty due to a choice of $\alpha_s$, 
this result indicates that the lattice spacing dependence
of the $B$ meson decay constant is sizably reduced after including 
the $O(\alpha_s a)$ and $O(\alpha_s/M)$ mixing terms.

\subsection{Estimate of systematic errors}
\label{sec:Systematic_Errors}

We now discuss possible sources of systematic errors
and attempt an order estimate of their magnitude.

As already discussed the uncertainty from the scale 
for the strong coupling constant, which is an $O(\alpha_s^2)$ effect, 
is quite significant. 
The magnitude of this error, estimated as half the difference of values 
for $q^{\ast}=\pi/a$ and $1/a$ is given in Table~\ref{tab:systematic_errors}
for each $\beta$.

We employ a light quark action which is $O(a)$-improved
at one-loop level.  Since the two-loop uncertainty in this 
improvement of $O(\alpha_s^2 a\LmQCD)$ is negligibly small, 
we expect the leading discretization error from the light 
quark sector to be $O((a\LmQCD)^2)$, which is also the magnitude 
of scaling violation in the gluon sector.
With a nominal value $\LmQCD$= 300~MeV, we 
estimate its size to be 2\%--8\% depending on $\beta$ as listed
in Table~\ref{tab:systematic_errors}.

Our results are obtained for NRQCD-I which represents the 
leading term in an expansion in $1/M$.  
We examine possible corrections due to the truncation 
by explicitly comparing the results of NRQCD-I with those of 
NRQCD-II which is correct to $O(1/M^2)$. 
Figure~\ref{fig:NRQCD-I_vs_-II} shows this comparison. 
We find that the $1/M^2$ correction does not exceed the statistical error, 
which is about 4\% in the $B$ meson mass region, as previously
observed in Ref.~\cite{HIROSHIMA}. 
Higher order uncertainties are expected to be even smaller. 

Another source of the systematic error is the perturbative matching of 
the action and the operators of NRQCD. 
In the one-loop calculation of the self-energy and the
current renormalization, we have consistently included all
terms of order $1/M$.
Hence $O(\alpha_s/(aM))$ corrections are properly taken into account 
in our calculation, and the leading error is of $O(\alpha_s/(aM)^2)$.
An order estimate for $O(\alpha_s/(aM)^2)$ is
given in Table~\ref{tab:systematic_errors} for each
$\beta$.
The magnitude increases for larger $\beta$ since $aM$ becomes 
smaller. 

Adding these four leading systematic errors in quadrature,
we find the systematic error to be about 8\% at $\beta$=6.1 and 5.9,
while it is much larger ($\sim$ 15\%) at $\beta$=5.7.
We, therefore, take an average over the two results at
$\beta$=6.1 and 5.9 to quote our final result.
In doing so, we have confirmed that the difference between
6.1 and 5.9 is within the estimated systematic error, as it
should be.

In addition to the above systematic uncertainties,
we must include an uncertainty in the lattice
scale $1/a$. 
Throughout this work we have used the scale set with the
string tension $\sqrt{\sigma}$.
Taking a variation of the ratio $m_{\rho}/\sqrt{\sigma}$ over 
$\beta=5.9, 6.1$ and 6.3, we assign a 3.5\% error in
the lattice scale as we did in Ref.~\cite{JLQCD}.

\subsection{Results}
\label{sec:Results}

Our final result for the $B$ meson decay constant in the quenched
approximation is given by
\begin{eqnarray}
  f_B       &=& 167(7)(15)\mbox{\ MeV},\\
  f_{B_{s}} &=& 191(4)(17)(^{+4}_{-0})\mbox{\ MeV}.
\end{eqnarray}
Here the central value is an average over the results at
$\beta$=6.1 and 5.9, and the errors are statistical and
systematic in the given order.
The systematic error includes 8\% 
as estimated in the previous subsection and the error in the 
lattice scale of 3.5\%,  added in quadrature.
For $f_{B_{s}}$ there is an additional uncertainty
from the strange quark mass.
We take the value from the $K$ meson mass ($\kappa_{s1}$)
for our central value.  Employing the $\phi$ mass ($\kappa_{s2}$) 
gives a larger $f_{B_{s}}$, which is given in the final
error. 

Our result is higher than that of Ali~Khan 
\textit{et al.}~\cite{GLOK_paper} at $\beta$=6.0 
($f_B$ = 147(11)(16) MeV and $f_{B_{s}}$= 175(8)(18) MeV), though
consistent within one standard deviation.
To compare with the results obtained with the the $O(a)$-improved clover 
action, we quote from the studies of the Fermilab~\cite{FNAL} group 
and of JLQCD~\cite{JLQCD}, 
\begin{eqnarray}
  \label{eq:fB_relativistic}
  f_B 
  & = & 164 (^{+14}_{-11})(8) \mbox{\ MeV} \;\;
  \mbox{(Fermilab)}, \nonumber \\
  & = & 173 (4) (13) \mbox{\ MeV} \;\;
  \mbox{(JLQCD)}, \nonumber \\
  f_{B_s}
  & = & 185 (^{+13}_{-8})(9) \mbox{\ MeV} \;\;
  \mbox{(Fermilab)}, \nonumber \\
  & = & 199 (3) (14) \mbox{\ MeV} \;\;
  \mbox{(JLQCD)}. \nonumber
\end{eqnarray}
Our values with NRQCD are in good agreement with these
relativistic calculations.

\subsection{$f_{B_s}/f_B$}
\label{sec:f_B_s/f_B}

Many of systematic uncertainties that appear in the calculation 
of the pseudoscalar decay constant $f_{P_{(s)}}$ cancel,
if we consider the ratio $f_{P_s}/f_P$. 
In particular, the two-loop uncertainty in the matching of
the axial current cancels out explicitly.

Figure~\ref{fig:f_P_s/f_P_vs_1/M_P} presents the $1/M_P$
dependence of $f_{P_s}/f_P$. We observe only a mild 
$1/M_P$ dependence.
The difference between NRQCD-I and NRQCD-II is much smaller
than the statistical error, which shows that the contribution
of the $1/M^2$ terms is negligible.
Finally, plotting the ratio as a function of lattice spacing
(see Figure~\ref{fig:f_B_s/f_B_vs_a}), 
we find the results at three $\beta$ values to be consistent 
with each other within the estimated errors.

Our result is
\begin{equation}
f_{B_{s}}/f_{B}=1.15(3)(1)(^{+3}_{-0}),
\end{equation}
which is obtained by averaging data at $\beta$=6.1 and
5.9.  The errors given are those from statistical, systematic and 
uncertainty in $\kappa_s$. 
For the systematic error the leading contribution arises from
the lack of the strange quark mass dependent renormalization,
which is $O(\alpha_s a\LmQCD)$ for the deviation $f_{B_{s}}/f_{B}-1$.
The magnitude is 3-5\% assuming the order counting.

\section{Mass Splittings}
\label{sec:Mass_Splittings}

A byproduct of our simulation is the mass difference 
between the $B$ and $B_s$ mesons, which can be compared with 
experiment.
Since the heavy quark mass cancels in this difference, there 
are no direct perturbative corrections for this quantity, though  
they enter implicitly through determination of the bare $b$-quark mass.

We plot the $1/M_P$ dependence of the $B_s-B$ mass difference in
Figure~\ref{fig:M_P_s-M_P_vs_1/M_P}, where we observe the dependence 
to be small.
The lattice spacing dependence is shown in 
Figure~\ref{fig:M_B_s-M_B_vs_a}.  A variation of about 20\%, 
beyond the statistical error of 8\%, is seen between the data at
$\beta$=6.1 and 5.9, which may represent scaling violation effects. 
Our result from an average over $\beta$=6.1 and 5.9 is given by
\begin{equation}
  M_{B_{s}}-M_B =
  87(7)(4)(^{+19}_{-0}) \mbox{\ MeV},
\end{equation}
where, as before, the errors represent statistical, systematic 
and $\kappa_s$
uncertainty. 
The possible systematic error is $O((a\LmQCD)^2)$, which is 2-3\%,
and the uncertainty of $1/a$$\sim$ 3.5\%. We estimate the error at 5\%
adding them in quadrature.
The dominant error is the uncertainty of $\kappa_s$.
It is encouraging that our result agrees with experimental value
$90 \pm 2$ MeV.

The hyperfine splitting $M_{B^*}-M_B$ is another
experimentally measured quantity.
Previous lattice studies in the quenched approximation 
have found that 
the hyperfine splitting of heavy-light and heavy-heavy mesons  
are much smaller than experiment~\cite{Mass_Splittings_Review}.
A possible reason for this discrepancy is an inappropriate value of the 
coupling $c_1$ for the $g\bbox{\sigma}\cdot\bbox{B}/2M_{0}$ term, for
which we use the tadpole improved tree-level value.
Since the hyperfine splitting of heavy-light
mesons is proportional to $c_1$, and that of heavy-heavy mesons to $c_1^2$,
it is possible that large corrections of $O(\alpha_s)$ remain 
(the non-perturbative calculation of this coupling has been done
 in Ref.~\cite{Trottier_Lepage_Perturbation_Lat97}, which reports
the possible $O(\alpha_s)$ correction).
Another possible source is the quenched approximation. 

The $1/M_P$ dependence of the hyperfine splitting obtained
in our simulation is shown in Figure~\ref{fig:hyperfine_vs_1/M_P}.
We observe that in the static limit the splitting linearly
vanishes due to the heavy quark symmetry.
Figure~\ref{fig:htperfine_vs_a} shows the lattice spacing dependence
of the splitting together with the experimental value of $M_{B^*}-M_B = 45.8 \pm 0.4$ MeV. 
While scaling is reasonably satisfied with our results, 
the magnitude is far below experiment. 
Averaging the two values at $\beta$=6.1 and 5.9, we find 
\begin{eqnarray}
  M_{B^*}-M_B       &=& 23(7)(5) \mbox{\ MeV},\\
  M_{B_s^*}-M_{B_s} &=& 26(3)(5)(^{+1}_{-0}) \mbox{\ MeV},
\end{eqnarray}
where we assume a 20\% systematic error since
the $O(\alpha_s)$ correction is not known for $c_1$.

\section{Conclusions}
\label{sec:Conclusions}

In this article we have presented a scaling study of the heavy-light
meson decay constant using lattice NRQCD with
the heavy-light current improved to $O(\alpha_s a)$ and
to $O(\alpha_s/M)$ consistently to the one-loop order in perturbation 
theory taking into account mixings with the relevant higher dimensional 
operators.
We have found the effect of the improvement to be substantial: 
the large $a$ dependence of $f_B$ is almost removed by the improvement.
This is most apparent in the static limit, where the effect is purely 
$O(\alpha_s a)$, but a similar
effect is also seen for the physical $B$ mass.

The two main sources of systematic errors in our results are 
perturbative two-loop corrections, $O(\alpha_s^2)$, in the renormalization 
factors for the NRQCD action, $O(a)$-improved Wilson action and
the axial vector current, and the one-loop corrections of 
$O(\alpha_s/(aM)^2)$ in the coefficients of the NRQCD action and the axial 
vector current.  The former uncertainty is quite large at $\beta=5.7$,
but diminishes to a 5\% level at weaker 
couplings of $\beta=5.9$ and 6.1.  The latter error, on the other hand,
increases toward smaller lattice spacings, reaching $\sim$ 6\%
at $\beta=6.1$. This counter increase of error represents
the limitation of lattice NRQCD itself;
the method breaks down once the heavy quark mass becomes 
smaller than the inverse lattice spacing.

The validity of a lattice NRQCD calculation of $f_B$ 
hinges on the existence of a window in lattice spacing over which 
the two errors as well as scaling violations are small.
We find these conditions to be satisfied at 
$\beta=5.9-6.1$ where the combined systematic errors are estimated to be
8\% (including the truncation in the $1/M$ expansion).
Achieving better accuracy with NRQCD would require
two-loop calculations to extend the window toward larger lattice spacings 
where the $O(\alpha_s/(aM)^2)$ error will become smaller. 
 
Another method for calculating heavy quark quantities on the lattice 
is the non-relativistic interpretation of relativistic actions~\cite{EKM}.
The advantage is that a continuum extrapolation 
can be carried out. The simulations by Refs.~\cite{JLQCD,FNAL,MILC}
have shown that the $a$ dependence in the heavy-light decay constant
is small for currently accessible range of $\beta=5.7\sim6.3$. 
A continuum extrapolation, with either constant or linear fit in 
the lattice spacing, has yielded the decay constants with
a systematic error of about 10\%.
A subtle point with this method, however, is that the $a$ 
dependence of systematic errors is non-linear.  Hence, strictly
speaking, it is not correct to extrapolate the
result with a simple linear or quadratic function of $a$.
To achieve a prediction of the $B$ meson decay constant more
accurate than is available, one needs 
to improve the action and currents so that the systematic
errors at finite values of $\beta$ are further reduced.
In this sense the study of $O(\alpha_s a)$ improvement
should be necessary. 

In spite of the limitations still inherent in the present 
results with NRQCD and with relativistic actions, 
we find it encouraging that the efforts with the two approaches 
have now yielded predictions for the $B$ meson decay constant 
in mutual agreement within the total error of about 10\% for 
quenched QCD. 

\section*{Acknowledgment}

This work is supported by the Supercomputer Project No.32
(FY1998) of High Energy Accelerator Research Organization (KEK),
and also in part by the Grants-in-Aid of the Ministry of Education
(Nos. 08640404, 09304029, 10640246, 10640248, 10740107, 10740125).
K-I.I, S.K., H.M. and S.T. are supported by the JSPS Research Fellowship.

\newcommand{\figwidth}{3.2in}
\newcommand{\figleftmarg}{-5mm}
\newcommand{\figsep}{-4mm}

\begin{figure}
  \begin{center}
    \vspace*{\figsep}
    \leavevmode\hspace{\figleftmarg}
    \psfig{file=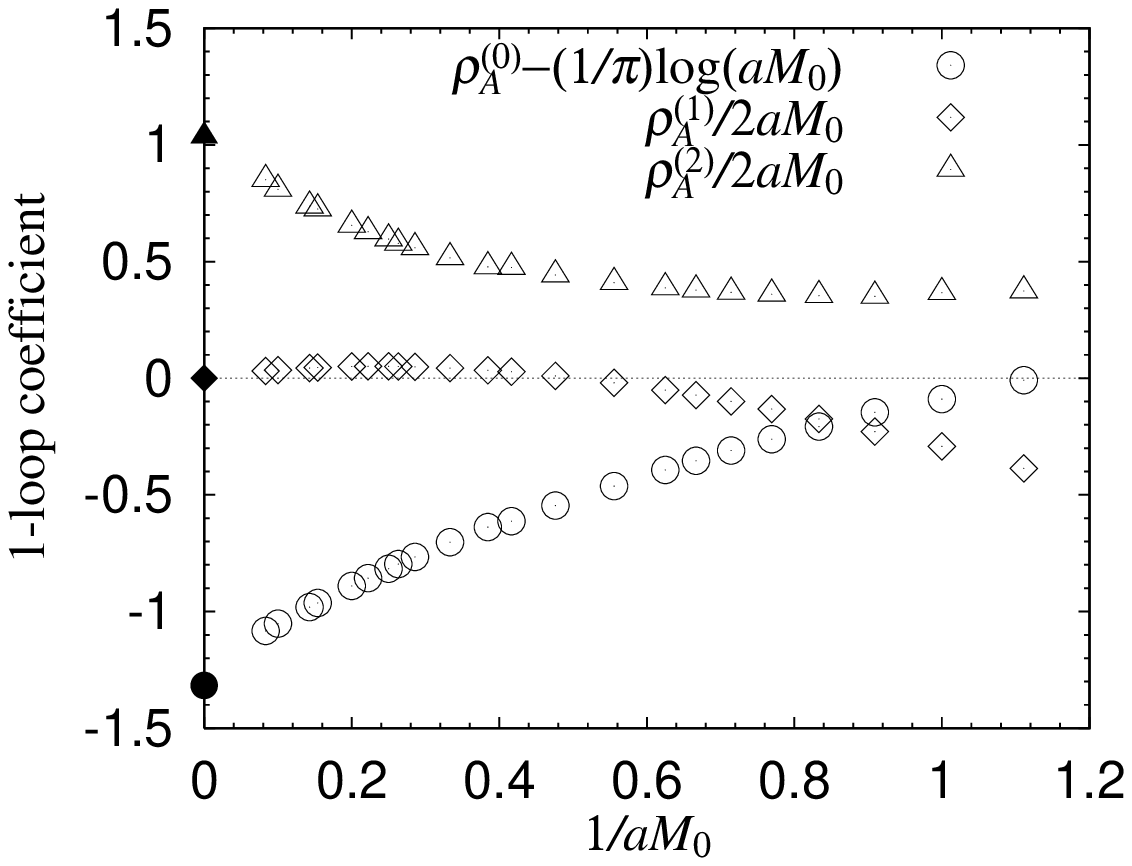,width=\figwidth}
  \end{center}
  \caption{$1/aM_{0}$ dependence of the one-loop
    coefficients for the axial vector current.
    Circles represent $\rho_A^{(0)}-1/\pi\ln(aM_0)$.
    Diamonds and triangles are $\rho_A^{(1)}/2aM_0$ and
    $\rho_A^{(2)}/2aM_0$ respectively.
    The static limit is shown with the filled symbols.
    }
  \label{fig:one-loop_coefficients}
\end{figure}

\clearpage
\begin{figure}
  \begin{multicols}{2}
    \begin{center}
      \vspace*{\figsep}
      \leavevmode\hspace{\figleftmarg}
      \psfig{file=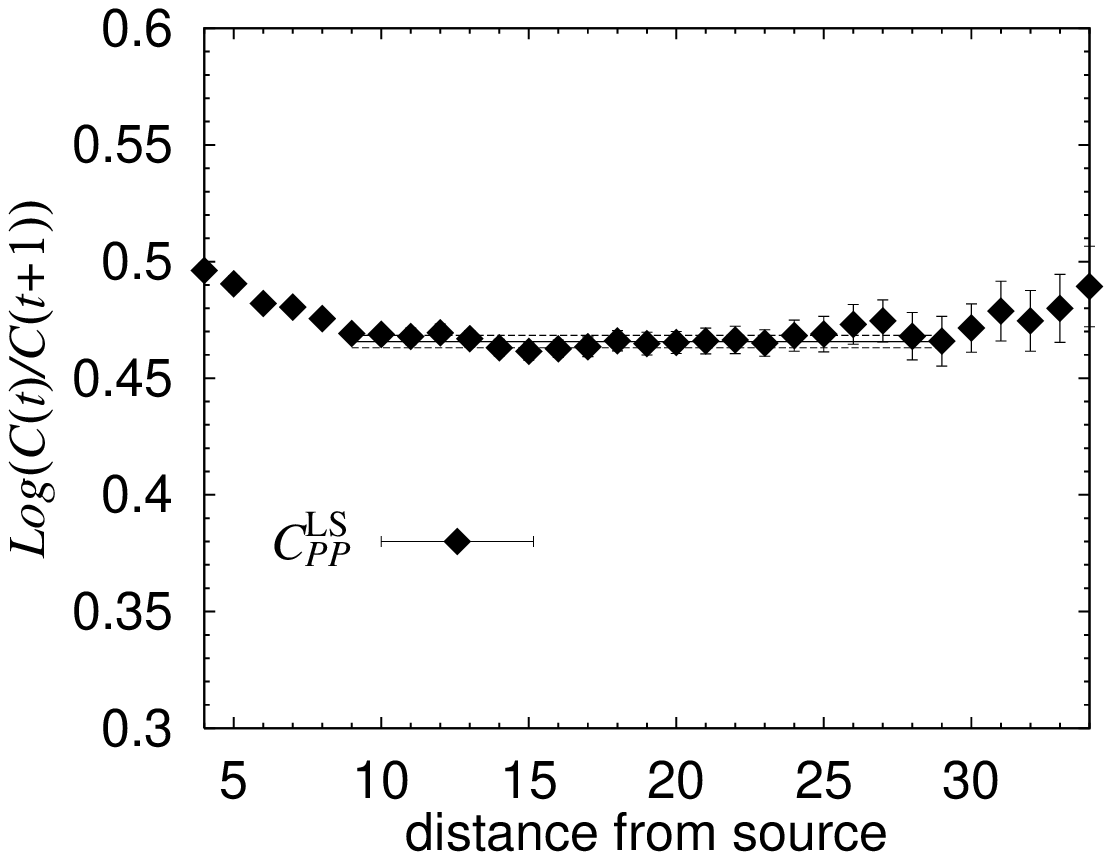,width=\figwidth}
    \end{center}
    \begin{center}
      \vspace*{\figsep}
      \leavevmode\hspace{\figleftmarg}
      \psfig{file=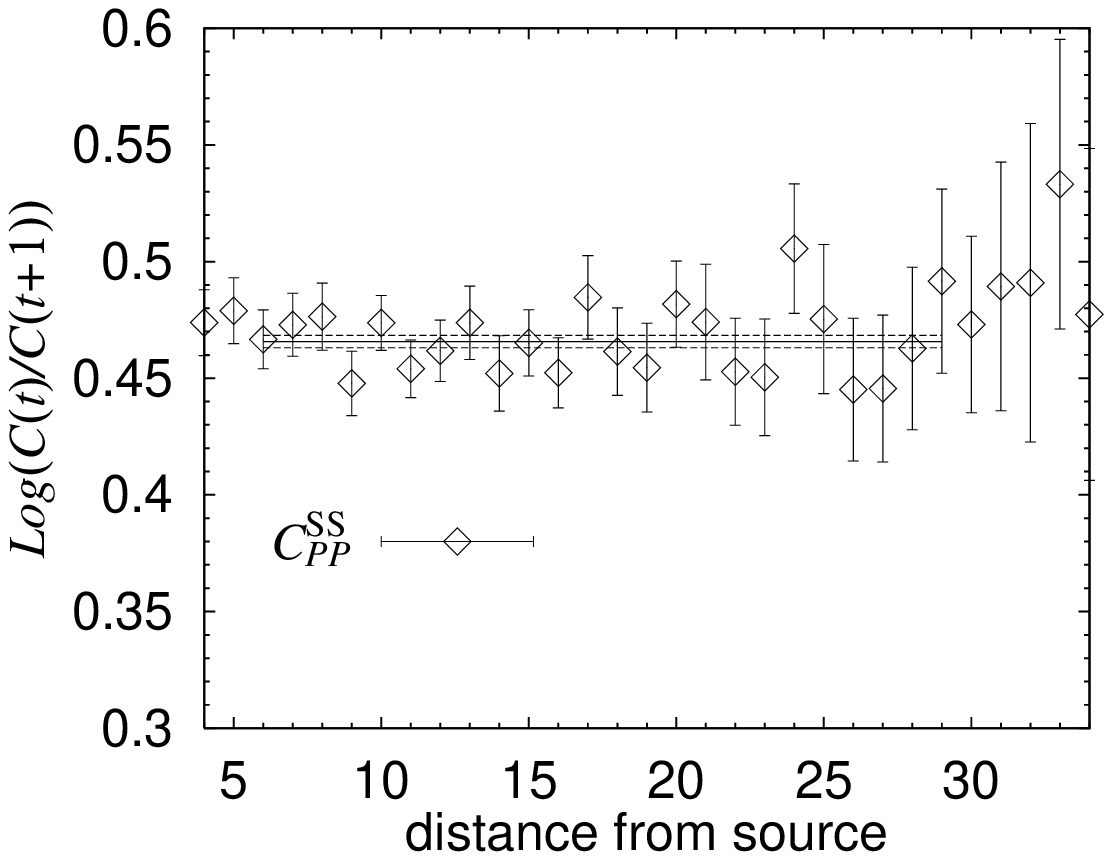,width=\figwidth}
    \end{center}
    \hfill
    \begin{center}
      \vspace*{\figsep}
      \leavevmode\hspace{\figleftmarg}
      \psfig{file=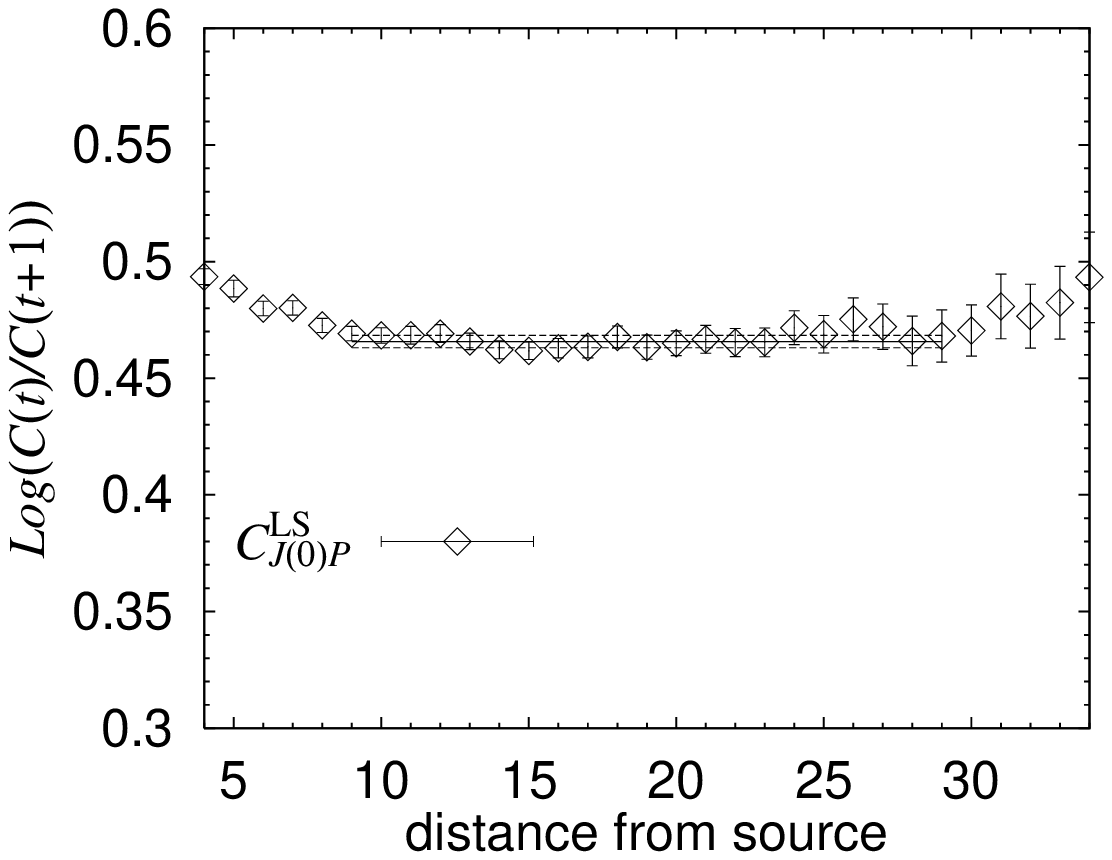,width=\figwidth}
    \end{center}
    \begin{center}
      \vspace*{\figsep}
      \leavevmode\hspace{\figleftmarg}
      \psfig{file=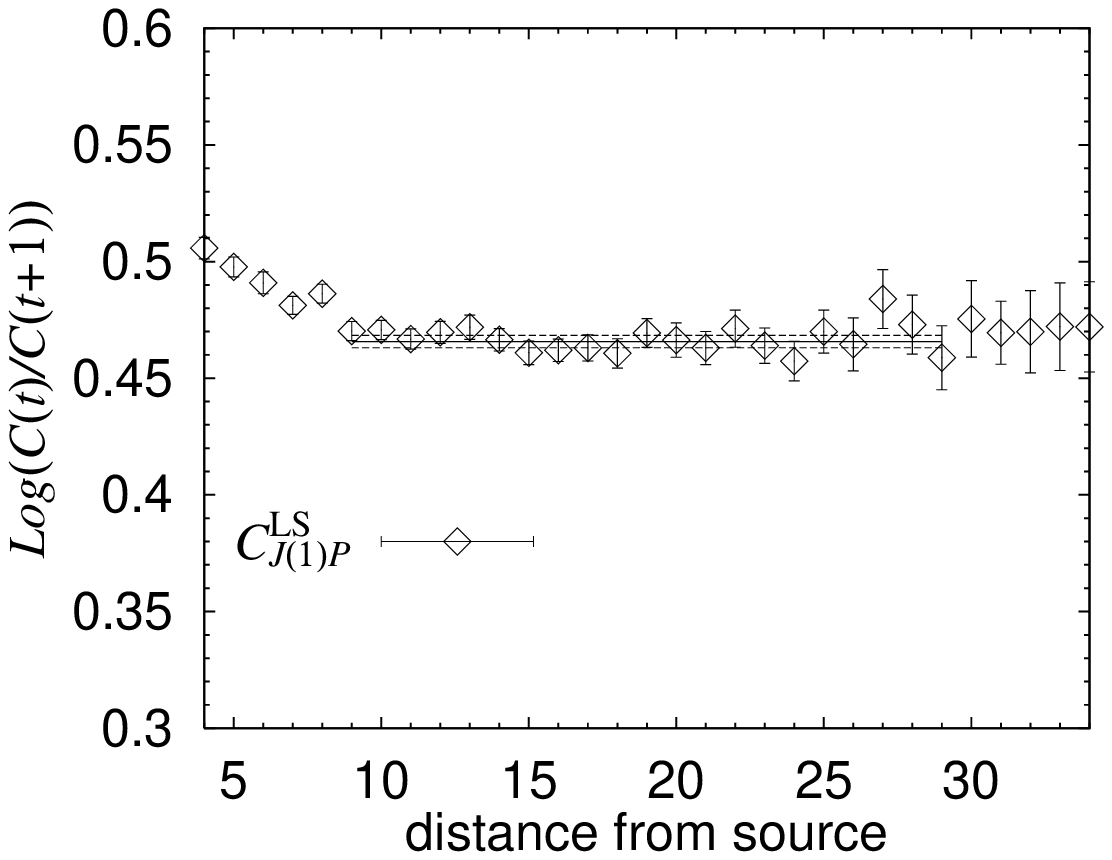,width=\figwidth}
    \end{center}
  \end{multicols}
  \caption{Effective mass of various correlators at
      $\beta$=6.1 and $(aM_0,n)=(2.1,2)$.
      The fitted value of $aE^{\mbox{\scriptsize bin}}$ is 
      shown by a solid line, and the error is indicated by
      dashed lines.
      The light quark hopping parameter $\kappa$=0.13586 is
      our heaviest one.}
  \label{fig:effective_mass_heaviest}
\end{figure}

\clearpage
\begin{figure}
  \begin{multicols}{2}
    \begin{center}
      \vspace*{\figsep}
      \leavevmode\hspace{\figleftmarg}
      \psfig{file=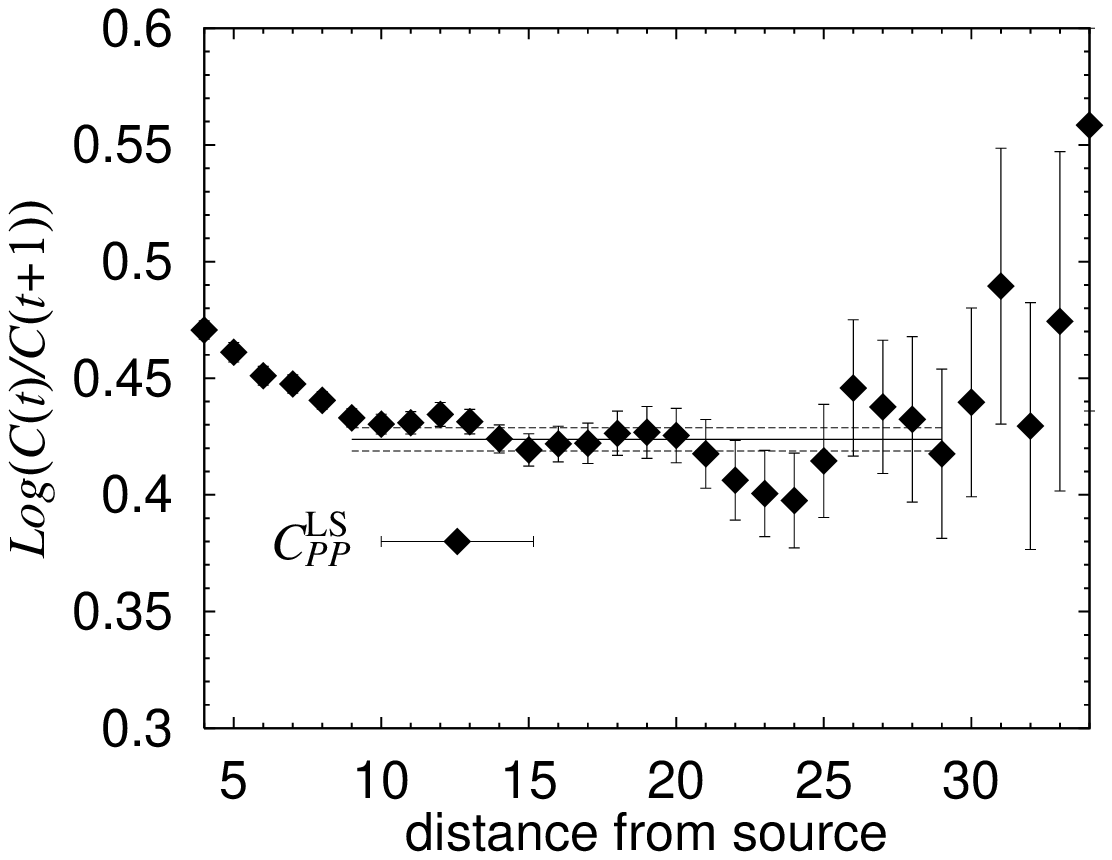,width=\figwidth}
    \end{center}
    \begin{center}
      \vspace*{\figsep}
      \leavevmode\hspace{\figleftmarg}
      \psfig{file=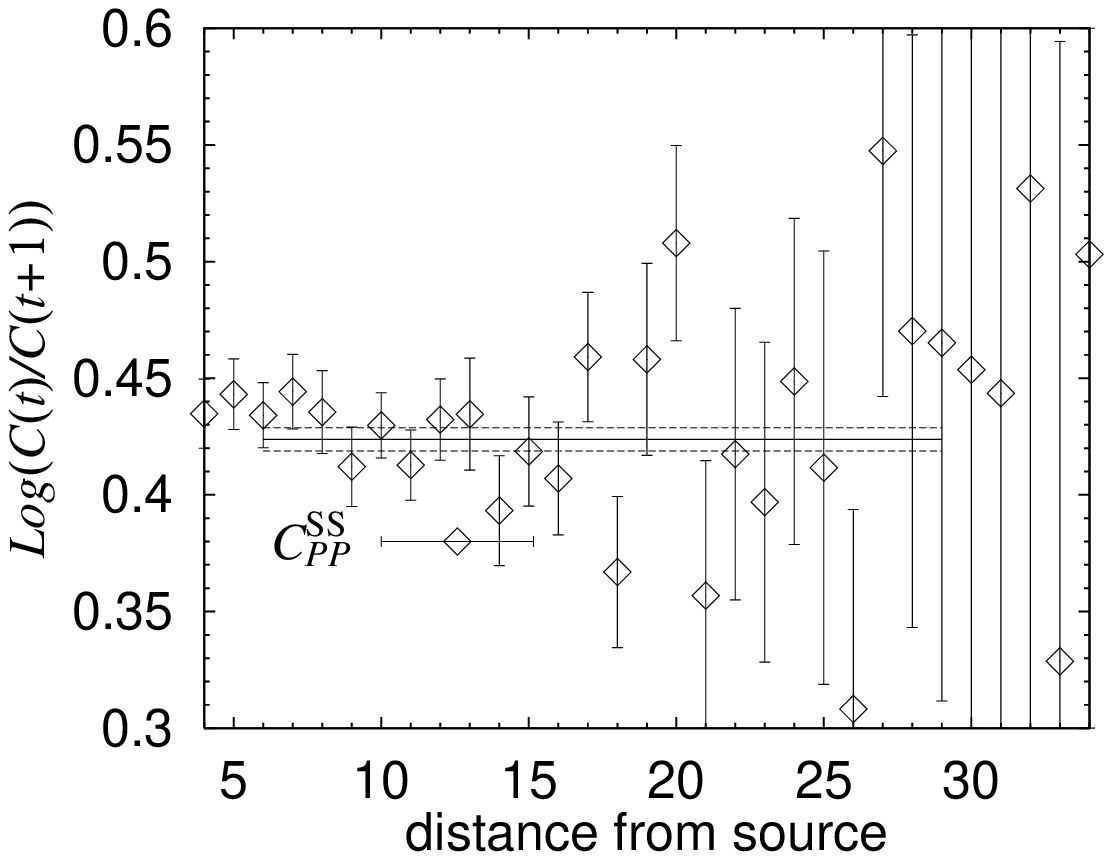,width=\figwidth}
    \end{center}
    \hfill
    \begin{center}
      \vspace*{\figsep}
      \leavevmode\hspace{\figleftmarg}
      \psfig{file=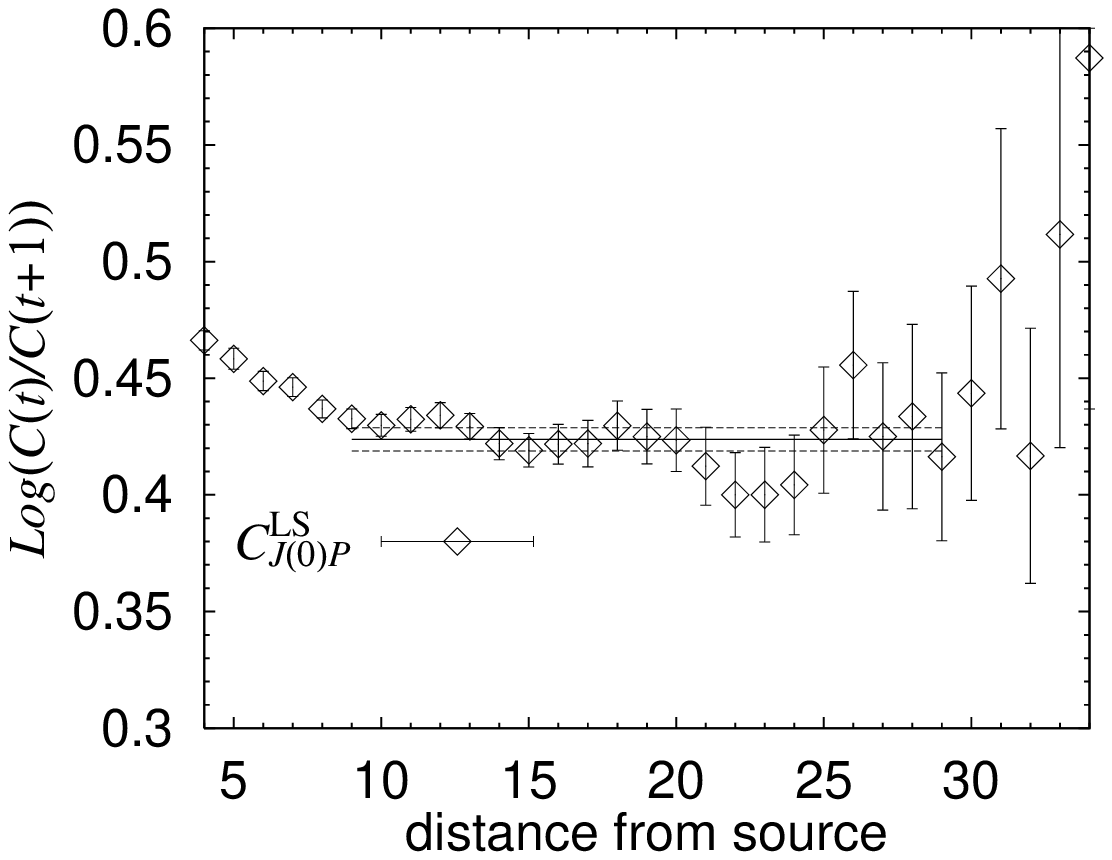,width=\figwidth}
    \end{center}
    \begin{center}
      \vspace*{\figsep}
      \leavevmode\hspace{\figleftmarg}
      \psfig{file=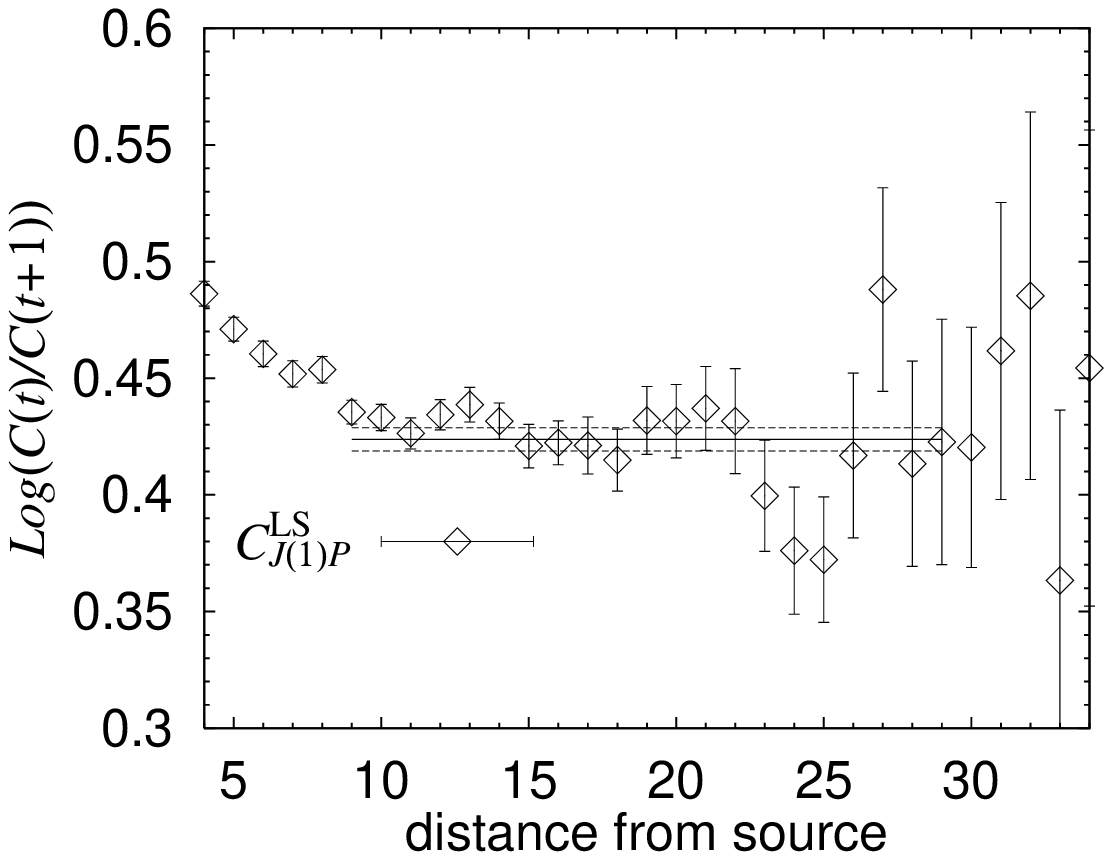,width=\figwidth}
    \end{center}
  \end{multicols}
  \caption{Same as Figure \ref{fig:effective_mass_heaviest},
    but with our lightest light quark mass $\kappa$=0.13716.}
  \label{fig:effective_mass_lightest}
\end{figure}

\clearpage
\begin{figure}
  \begin{center}
    \vspace*{\figsep}
    \leavevmode\hspace{\figleftmarg}
    \psfig{file=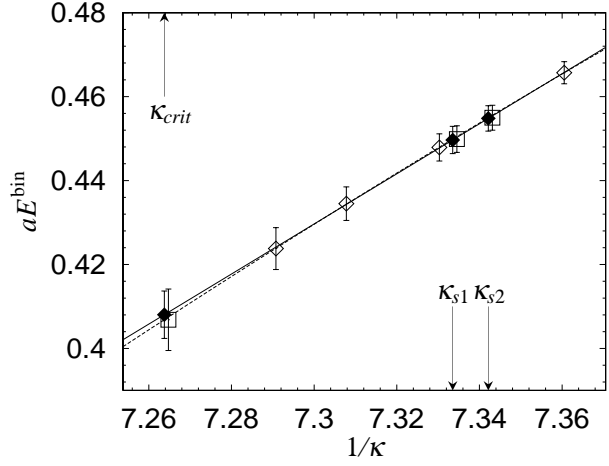,width=\figwidth}
  \end{center}
  \caption{Chiral limit of the heavy-light binding energy
    $aE^{\mbox{\scriptsize bin}}$ at $\beta$=6.1
    and $(aM_{0}, n)=(2.1, 2)$.
    Open diamonds represent our data. Filled diamonds are
    the results in the chiral limit 
    ($\kappa_{\mbox{\scriptsize crit}}$)
    or in the strange quark mass ($\kappa_{s1}$ or $\kappa_{s2}$)
    with linear fitting (solid line), and open squares are
    the results with quatratic fitting (dotted line). }
  \label{fig:chiral_limit_binding_energy}
\end{figure}

\begin{figure}
  \begin{multicols}{2}
    \begin{center}
      \vspace*{\figsep}
      \leavevmode\hspace{\figleftmarg}
      \psfig{file=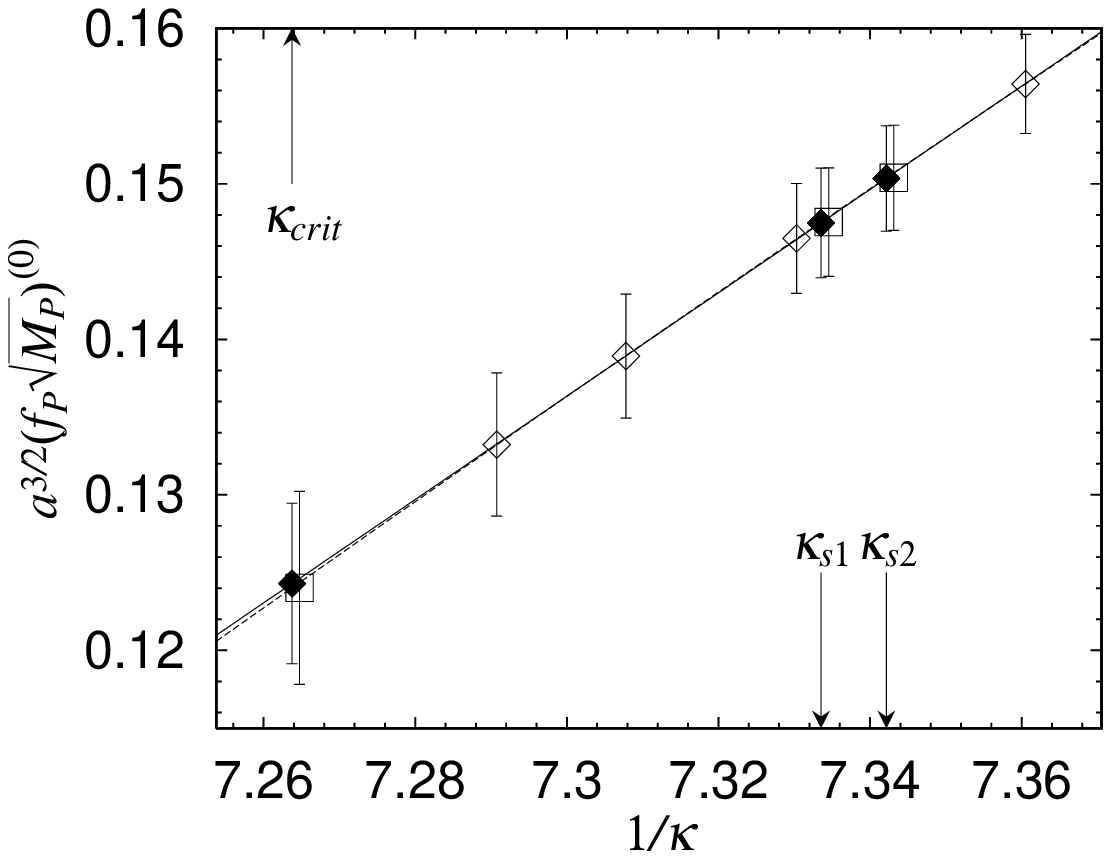,width=\figwidth}
    \end{center}
    \hfill
    \begin{center}
      \vspace*{\figsep}
      \leavevmode\hspace{\figleftmarg}
      \psfig{file=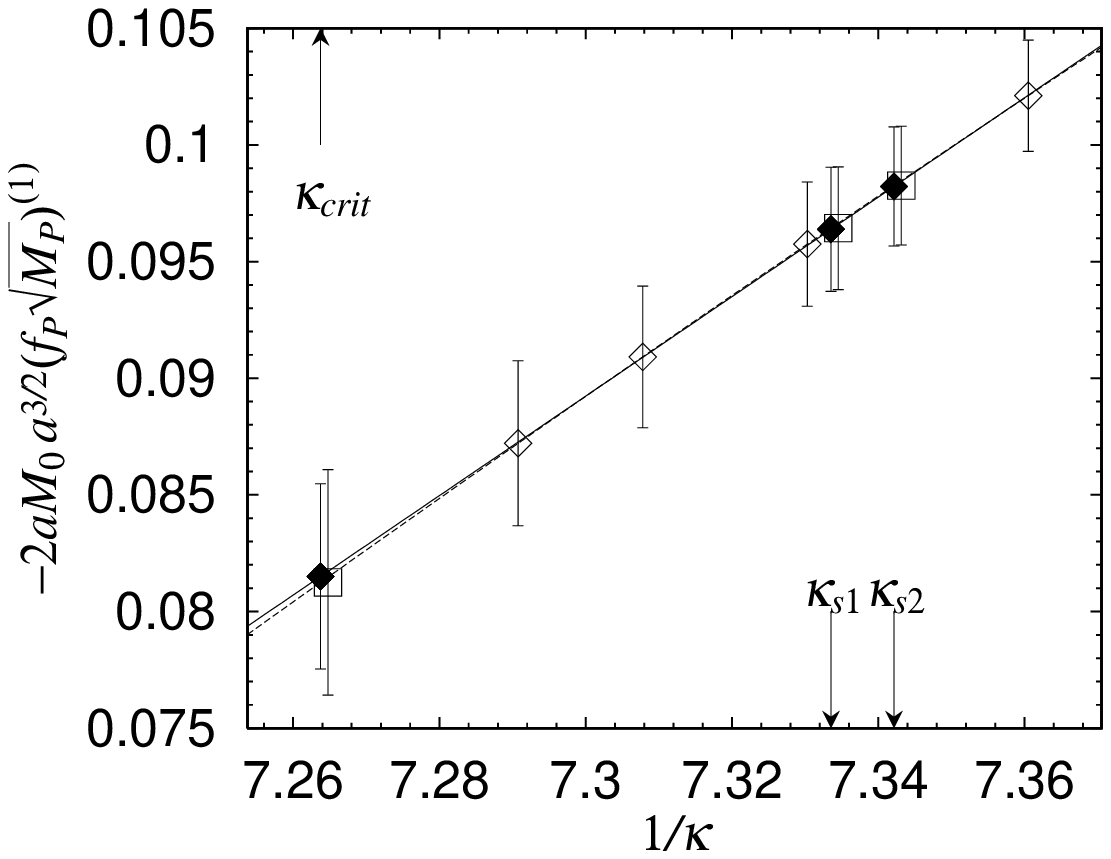,width=\figwidth}
    \end{center}
  \end{multicols}
  \caption{Chiral limit of the decay constant 
    $a^{3/2}(f_P\sqrt{M_P})^{(0)}$ (left) and 
    $-2aM_0 a^{3/2}(f_P\sqrt{M_P})^{(1)}$ (right) at
    $\beta$=6.1 and $(aM_0,n)=(2.1,2)$. 
    The meaning of the symbols is the same as that in
    Fig.~\ref{fig:chiral_limit_binding_energy}, }
  \label{fig:chiral_limit_decay_constant}
\end{figure}

\clearpage
\begin{figure}
  \begin{center}
    \vspace*{\figsep}
    \leavevmode\hspace*{\figleftmarg}
    \psfig{file=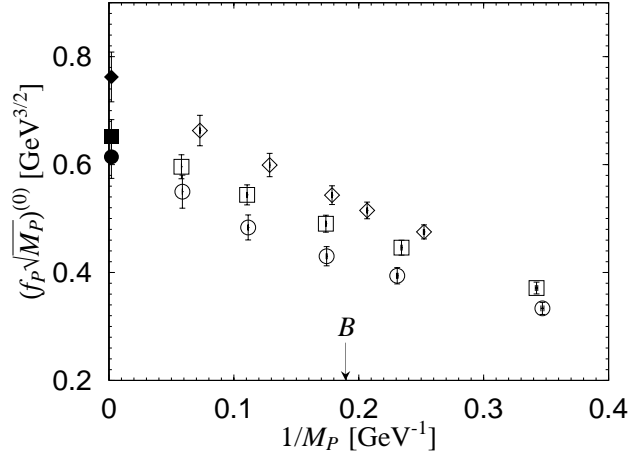,width=\figwidth}
  \end{center}
  \caption{$1/M_{P}$ dependence of $(f_P\sqrt{M_P})^{(0)}$.
    We used tree level value for $M_P$ in the plot.
    Data at three $\beta$ values are shown:
    $\beta$=5.7 (diamonds), 5.9 (squares), and 6.1
    (circles).
    The static limit (filled symbols) is obtained with a
    quadratic extrapolation.
    }
  \label{fig:fsqrtm_vs_1/M_P}
\end{figure}

\begin{figure}
  \begin{center}
    \vspace*{\figsep}
    \leavevmode\hspace*{\figleftmarg}
    \psfig{file=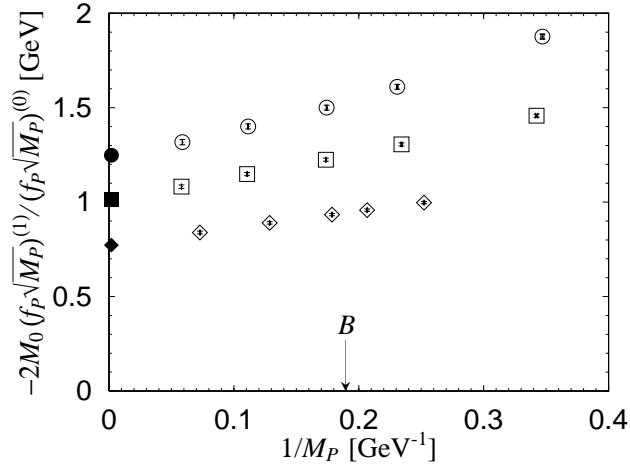,width=\figwidth}
  \end{center}
  \caption{Ratio of the leading and mixing operators 
    $-2M_0 (f_P\sqrt{M_P})^{(1)}/(f_P\sqrt{M_P})^{(0)}$.
    We used tree level results for $1/M_P$ in the plot.
    Data at three $\beta$ values are shown:
    $\beta$=5.7 (diamonds), 5.9 (squares), and 6.1
    (circles).
    The static limit (filled symbols) is obtained with a
    quadratic extrapolation.
    }
  \label{fig:fsqrtm^1/fsqrtm^0}
\end{figure}

\clearpage
\begin{figure}
  \begin{multicols}{2}
    \begin{center}
      \vspace*{\figsep}
      \leavevmode\hspace{\figleftmarg}
      \psfig{file=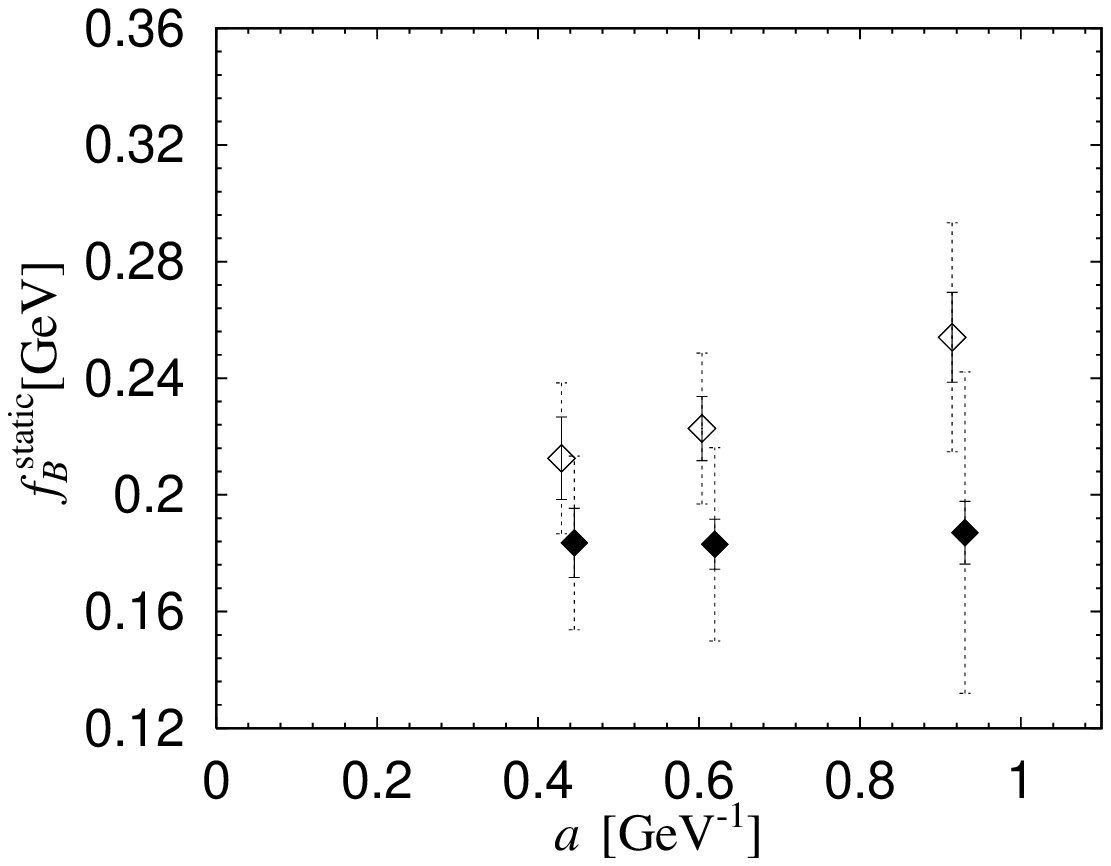,width=\figwidth}
    \end{center}
    \hfill
    \begin{center}
      \vspace*{\figsep}
      \leavevmode\hspace{\figleftmarg}
      \psfig{file=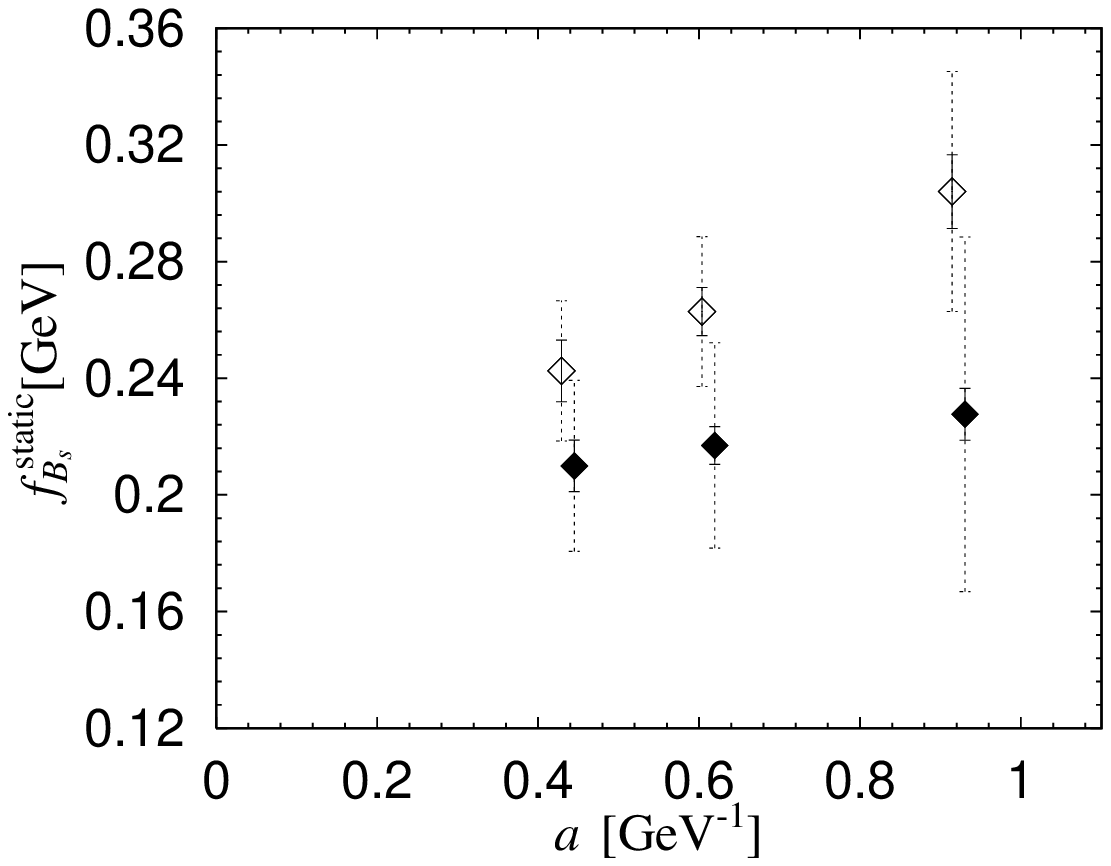,width=\figwidth}
    \end{center}
  \end{multicols}
  \caption{The lattice spacing dependence of
    $f_B^{static}$ at $\kappa=\kappa_{\mbox{\scriptsize crit}}$ (left) and
    $\kappa=\kappa_{s1}$ (right). 
    Open diamonds represent the result without the operator
    mixing, while filled diamonds include the mixing effect.
    The symbols show the $q^*$ averaged results, and
    are slightly shifted in horizontal axis so that error bars do not overlap.
    Solid error bars show the statistical error, and
    dashed ones show the uncertainty of $q^*$ 
    from the difference of the two choices of $q^*=\pi/a$ and $1/a$.
    }
  \label{fig:fsqrtm_static}
\end{figure}

\clearpage
\begin{figure}
  \begin{center}
    \vspace*{\figsep}
    \leavevmode\hspace*{\figleftmarg}
    \psfig{file=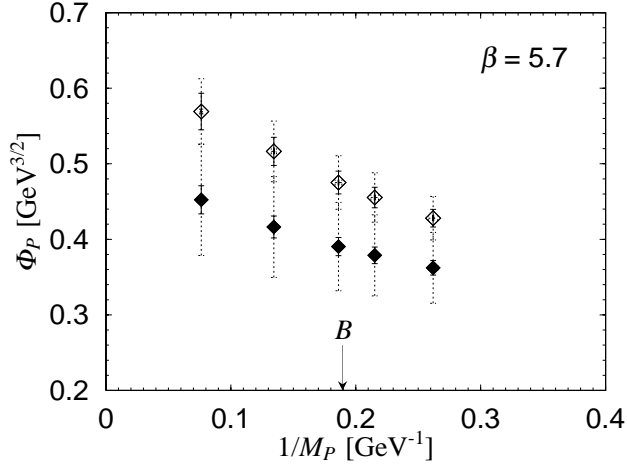,width=\figwidth}
  \end{center}
  \begin{center}
    \vspace*{\figsep}
    \leavevmode\hspace*{\figleftmarg}
    \psfig{file=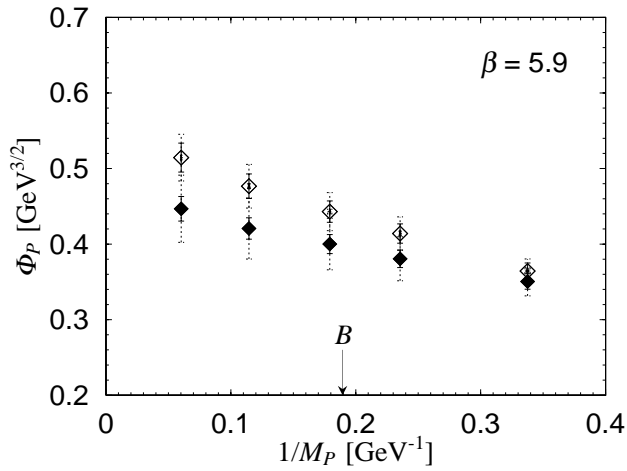,width=\figwidth}
  \end{center}
  \begin{center}
    \vspace*{\figsep}
    \leavevmode\hspace*{\figleftmarg}
    \psfig{file=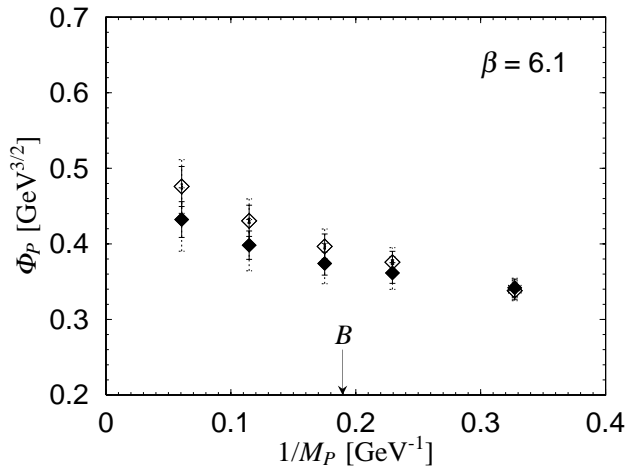,width=\figwidth}
  \end{center}
  \caption{$\Phi_P$ as a function of $1/M_P$ for each
    $\beta$. 
    Open diamonds represent the result without the operator
    mixing, while filled diamonds include the mixing effect.
    The symbols show the $q^*$ averaged results.
    Solid error bars show the statistical error, and
    dashed ones show the uncertainty of $q^*$.
    }
  \label{fig:Phi_vs_1/M_P}
\end{figure}

\clearpage
\begin{figure}
  \begin{multicols}{2}
    \begin{center}
      \vspace*{\figsep}
      \leavevmode\hspace{\figleftmarg}
      \psfig{file=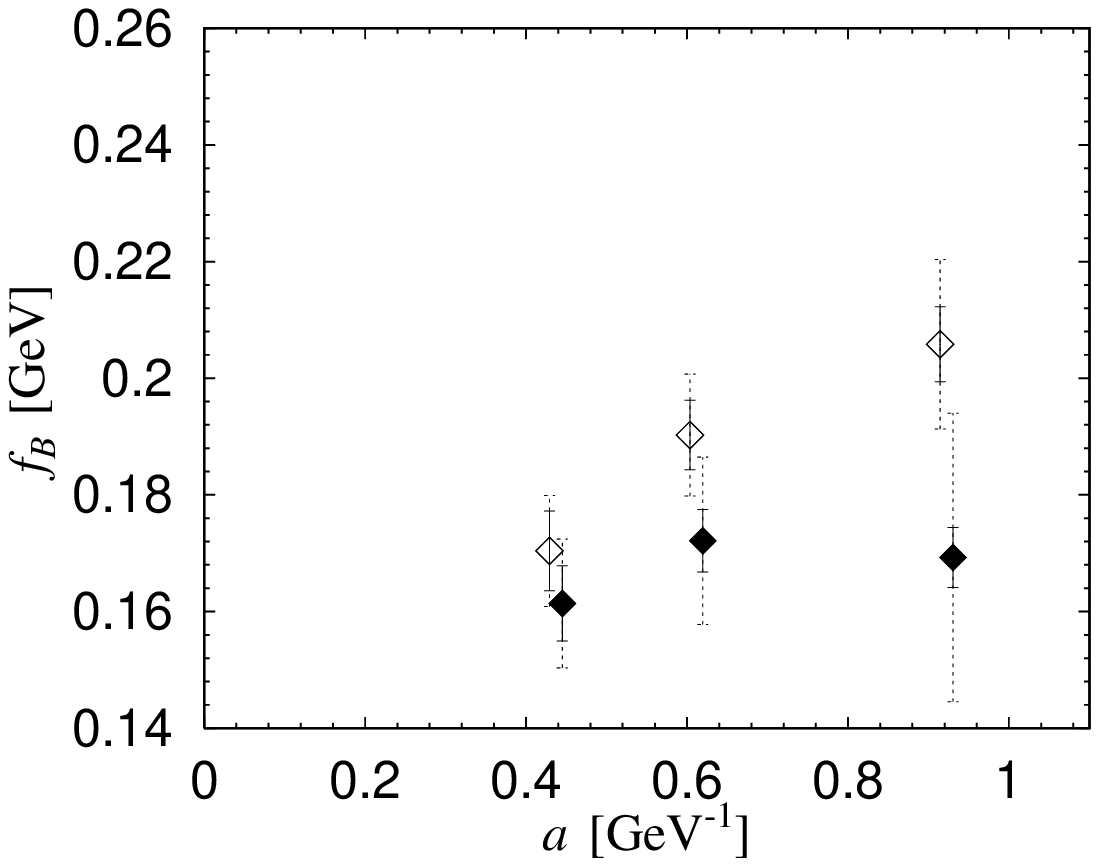,width=\figwidth}
    \end{center}
    \hfill
    \begin{center}
      \vspace*{\figsep}
      \leavevmode\hspace{\figleftmarg}
      \psfig{file=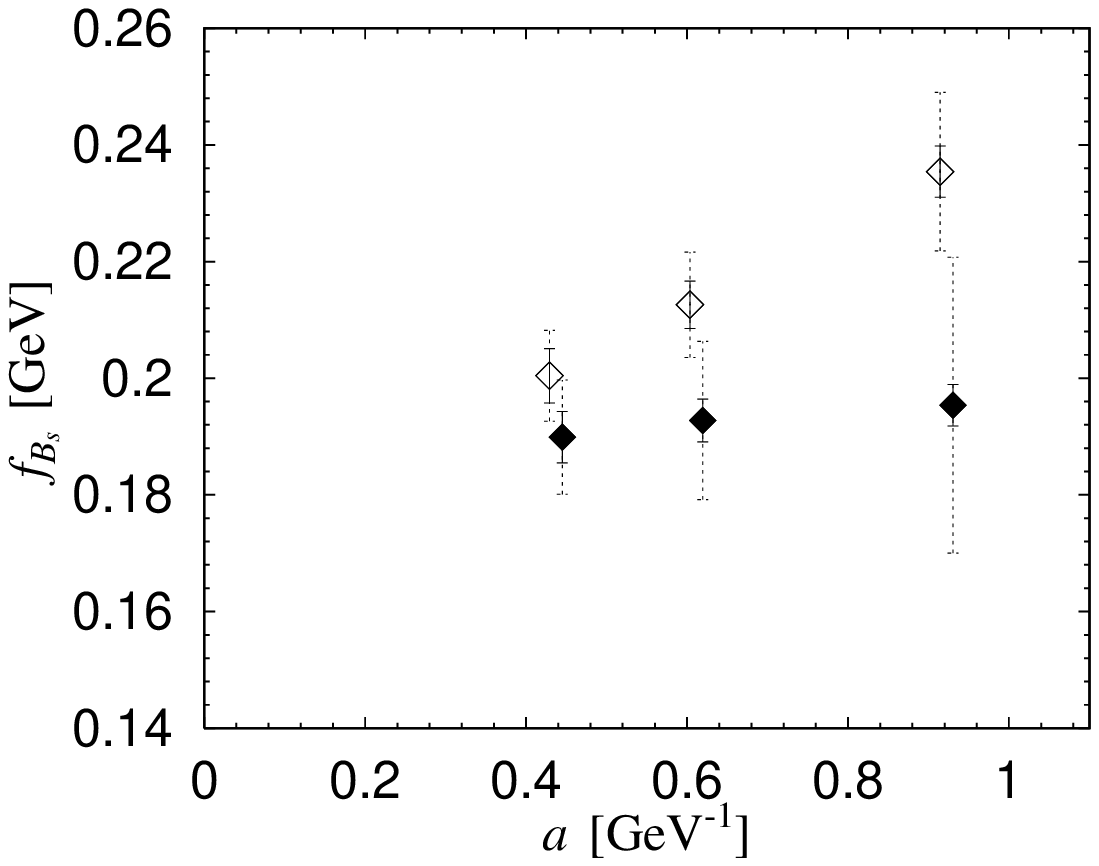,width=\figwidth}
    \end{center}
  \end{multicols}
  \caption{$a$ dependence of $f_{B}$ (left) and $f_{B_s} (right)$. 
    Filled symbols represent the result with the
    contribution from $J^{(1)}_{\mbox{\scriptsize latt}}$ 
    and $J^{(2)}_{\mbox{\scriptsize latt}}$.
    Open symbols do not include these effects.
    The symbols show the $q^*$ averaged results, and
    are slightly shifted in horizontal axis so that error bars do not overlap.
    Solid error bars show the statistical error, and
    dashed ones show the uncertainty of $q^*$. 
}
  \label{fig:f_B_vs_a}
\end{figure}

\begin{figure}
  \begin{center}
    \vspace*{\figsep}
    \leavevmode\hspace*{\figleftmarg}
    \psfig{file=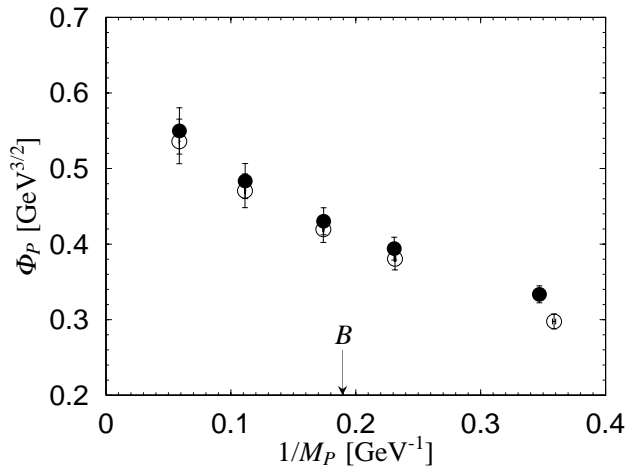,width=\figwidth}
  \end{center}
  \caption{Comparison of $\Phi_P$ from NRQCD-I (filled
    circles) with that from NRQCD-II (open circles) at tree level.
    }
  \label{fig:NRQCD-I_vs_-II}
\end{figure}

\clearpage
\begin{figure}
  \begin{center}
    \vspace*{\figsep}
    \leavevmode\hspace*{\figleftmarg}
    \psfig{file=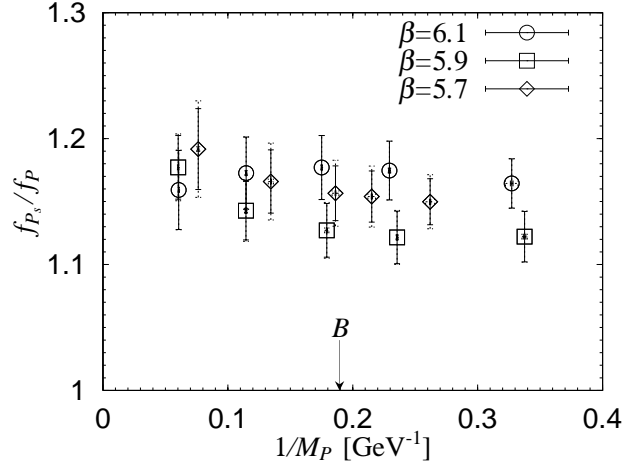,width=\figwidth}
  \end{center}
  \caption{1/$M_{P}$ dependence of $f_{P_{s}}/f_{P}$ 
    with $\kappa_{s1}$. The symbols show the $q^*$ averaged
    results. Solid error bars show the statistical error, and
    dotted ones show the uncertainty of $q^*$.}
  \label{fig:f_P_s/f_P_vs_1/M_P}
\end{figure}

\begin{figure}
  \begin{center}
    \vspace*{\figsep}
    \leavevmode\hspace*{\figleftmarg}
    \psfig{file=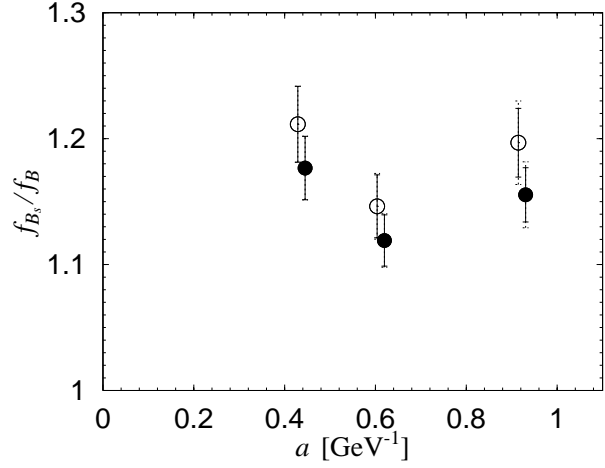,width=\figwidth}
  \end{center}
  \caption{$a$ dependence of $f_{B_s}/f_{B}$ with
    $\kappa_{s1}$ (filled circles) and with $\kappa_{s2}$
    (open circles).
    The symbols show the $q^*$ averaged
    results. Solid error bars show the statistical error, and
    dotted ones show the uncertainty of $q^*$.}
  \label{fig:f_B_s/f_B_vs_a}
\end{figure}

\clearpage
\begin{figure}
  \begin{center}
    \vspace*{\figsep}
    \leavevmode\hspace*{\figleftmarg}
    \psfig{file=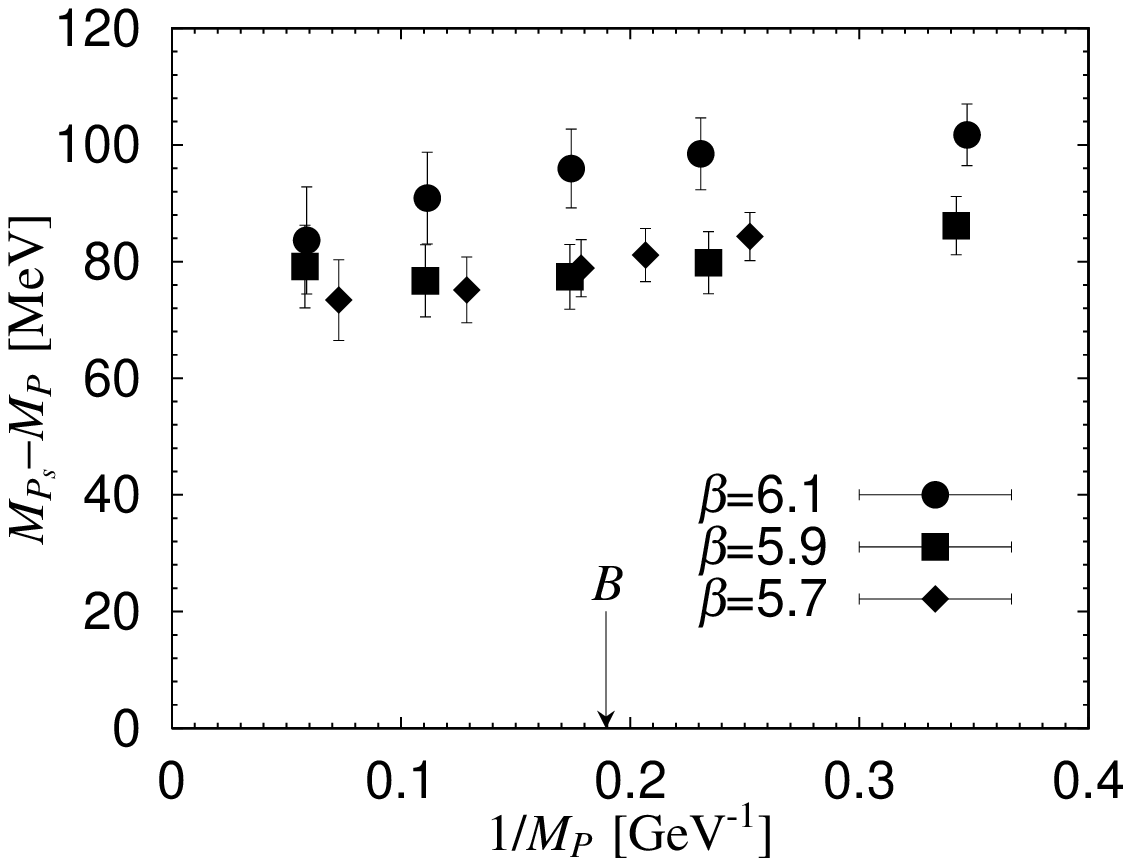,width=\figwidth}
  \end{center}
  \caption{$1/M_{P}$ dependence of $M_{P_{s}}-M_{P}$ 
    with $\kappa_{s1}$. We used the tree level results
    for $1/M_P$ in the plot.
  }
  \label{fig:M_P_s-M_P_vs_1/M_P}
\end{figure}

\begin{figure}
  \begin{center}
    \vspace*{\figsep}
    \leavevmode\hspace*{\figleftmarg}
    \psfig{file=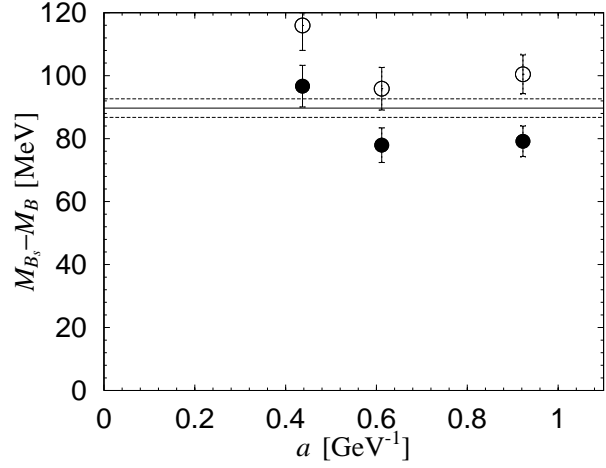,width=\figwidth}
  \end{center}
  \caption{$a$ dependence of $M_{B_s}-M_{B}$ with
    $\kappa_{s1}$ (filled circles) and with $\kappa_{s2}$
    (open circles).
    The experimental value is shown by a solid line.
    }
  \label{fig:M_B_s-M_B_vs_a}
\end{figure}

\clearpage
\begin{figure}
  \begin{center}
    \vspace*{\figsep}
    \leavevmode\hspace*{\figleftmarg}
    \psfig{file=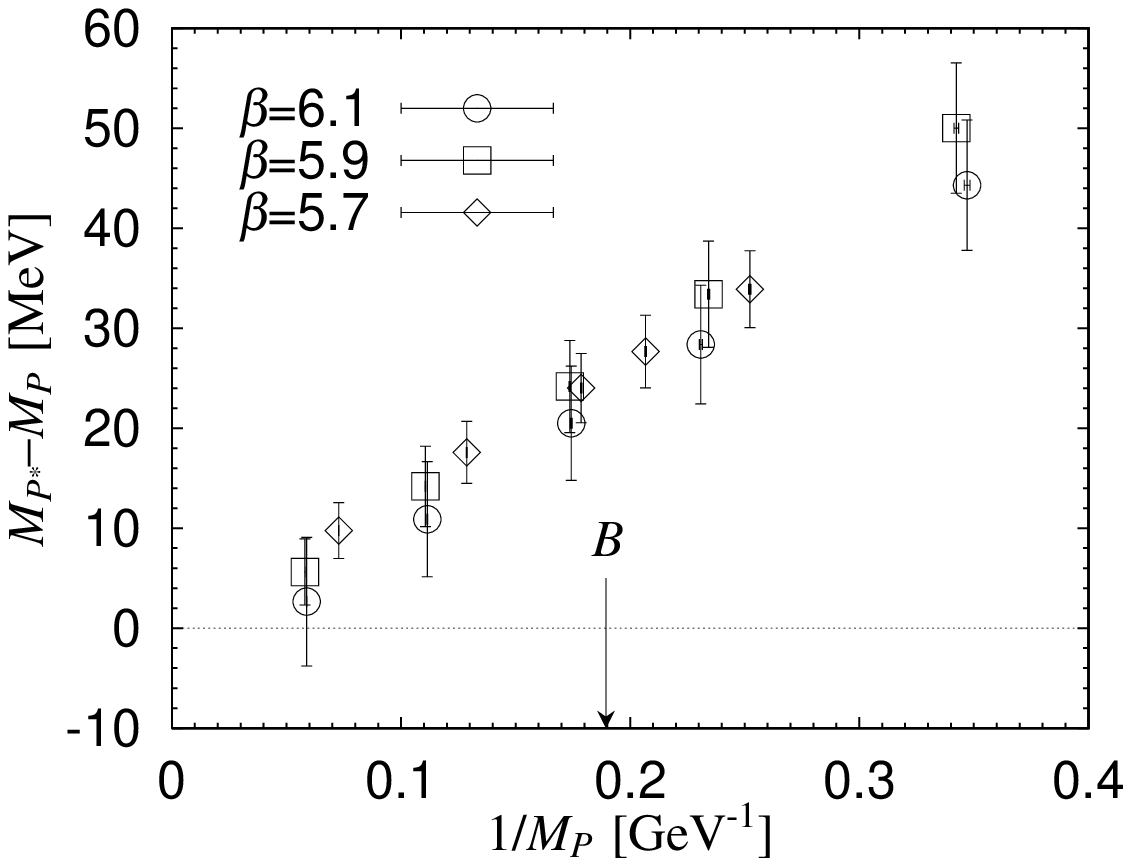,width=\figwidth}
  \end{center}
  \caption{$1/M_{P}$ dependence of $M_{P^*}-M_{P}$. We used
  the tree level results for $1/M_P$ in the plot.}
  \label{fig:hyperfine_vs_1/M_P}
\end{figure}

\begin{figure}
  \begin{center}
    \vspace*{\figsep}
    \leavevmode\hspace*{\figleftmarg}
    \psfig{file=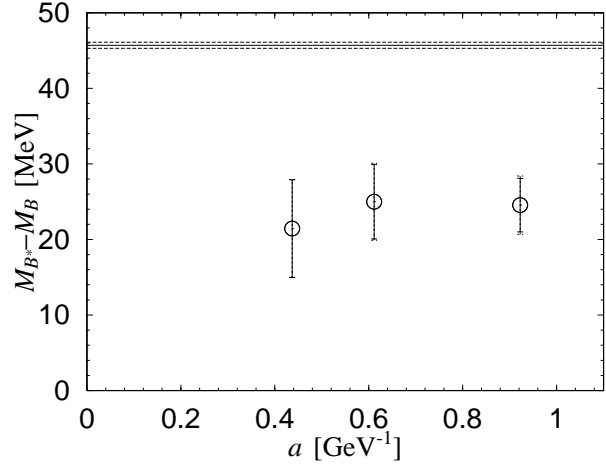,width=\figwidth}
  \end{center}
  \caption{$a$ dependence of $M_{B^*}-M_{B}$.
    The experimental value is shown by a solid line.
    }
  \label{fig:htperfine_vs_a}
\end{figure}
\begin{table}
  \begin{tabular}{cc|ccc|ccc}
    $aM_{0}$ & $n$ & $\rho^{(0)}_{A}-(1/\pi)\log(aM_{0})$ &
                     $\rho^{(1)}_{A}/2aM_{0}$ & $\rho^{(2)}_{A}/2aM_{0}$ &
    $A$ & $B$ & $C$ \\ \hline
    $\infty$
        &-&-1.317 & 0.000 & 1.036 & 1.069 & -     & 0.481 \\
    12.0&2&-1.162 & 0.026 & 0.851 & 1.022 &-0.091 & 0.312 \\
    10.0&2&-1.131 & 0.030 & 0.809 & 1.011 &-0.075 & 0.279 \\
     7.0&2&-1.061 & 0.036 & 0.737 & 0.983 &-0.032 & 0.197 \\
     6.5&2&-1.043 & 0.037 & 0.725 & 0.976 &-0.023 & 0.177 \\
     5.0&2&-0.970 & 0.040 & 0.656 & 0.946 & 0.018 & 0.094 \\
     4.5&2&-0.937 & 0.039 & 0.628 & 0.931 & 0.036 & 0.055 \\
     3.8&2&-0.876 & 0.036 & 0.578 & 0.903 & 0.069 &-0.014 \\
     3.5&2&-0.846 & 0.034 & 0.559 & 0.888 & 0.086 &-0.052 \\
     3.0&2&-0.782 & 0.026 & 0.516 & 0.855 & 0.119 &-0.127 \\
     2.1&3&-0.626 &-0.015 & 0.442 & 0.754 & 0.329 &-0.315 \\
     1.5&3&-0.433 &-0.108 & 0.378 & 0.621 & 0.456 &-0.542 \\
     1.3&3&-0.341 &-0.173 & 0.360 & 0.547 & 0.511 &-0.647 \\
     0.9&4&-0.088 &-0.445 & 0.374 & 0.300 & 0.782 &-0.921 
  \end{tabular}
  \caption{One-loop coefficients of the axial vector
    current $\rho_A^{(0)}$, $\rho_A^{(1)}$ and
    $\rho_A^{(2)}$ defined in Eq.~(\ref{eq:FULLA4}). 
    The self-energy corrections are also listed.}
  \label{tab:one-loop_coefficients}
\end{table}

\begin{table}
  \begin{tabular}{c|ccc}
    $\beta$  & 6.1     & 5.9     & 5.7 \\ \hline
    size & $24^{3}\times 64$ & $16^{3}\times 48$ & 
           $12^{3}\times 32$ \\ \hline
    \#conf   & 120     & 300     & 300     \\ \hline
    $c_{\mbox{\scriptsize sw}}$ 
             & 1.525   & 1.580   & 1.674   \\ \hline 
    $\kappa$ & 0.13586 & 0.13630 & 0.13690 \\
             & 0.13642 & 0.13711 & 0.13760 \\
             & 0.13684 & 0.13769 & 0.13840 \\
             & 0.13716 & 0.13816 & 0.13920 \\ \hline
    $u_{0}$  & 0.8816  & 0.8734  & 0.86087 \\ \hline
    $(aM_{0},n)$
             & (7.0,2) & (10.0,2)& (12.0,2) \\
             & (3.5,2) &  (5.0,2)&  (6.5,2) \\
             & (2.1,2) &  (3.0,2)&  (4.5,2) \\
             & (1.5,3) &  (2.1,3)&  (3.8,2) \\
             & (0.9,4) &  (1.3,3)&  (3.0,2) \\ \hline
    $\alpha_{V}(\pi/a)$ & 0.149 & 0.164 & 0.188 \\ \hline
    $\alpha_{V}(  1/a)$ & 0.229 & 0.270 & 0.355 \\ \hline
    $\kappa_{\mbox{\scriptsize  crit}}$ 
                   & 0.13767 & 0.13901 & 0.14157 \\ \hline
    $\kappa_{s1} $ & 0.13635 & 0.13702 & 0.13800 \\ \hline
    $\kappa_{s2} $ & 0.13609 & 0.13657 & 0.13707 \\ \hline
    $1/a$ (GeV)    & 2.29    & 1.64    & 1.08
  \end{tabular}
  \caption{Lattice parameters.}
  \label{tab:lattice_parameters}
\end{table}

\begin{table}
  \begin{tabular}{c|ccc|ccc|ccc}
    & \multicolumn{3}{c|}{$\kappa=\kappa_{\mbox{\scriptsize  crit}}$}
    & \multicolumn{3}{c|}{$\kappa=\kappa_{s1}$}
    & \multicolumn{3}{c}{ $\kappa=\kappa_{s2}$}\\
    \hline
    $aM_{0}$& $aE^{\mbox{\scriptsize bin}}$ & $q^{*}=\pi/a$ & $q^{*}=1/a$ &
              $aE^{\mbox{\scriptsize bin}}$ & $q^{*}=\pi/a$ & $q^{*}=1/a$ &
              $aE^{\mbox{\scriptsize bin}}$ & $q^{*}=\pi/a$ & $q^{*}=1/a$ \\
    \hline
    \multicolumn{10}{c}{$\beta$=5.7}\\ \hline
  12.0 &  0.669(11)& 12.271(11)& 11.919(11)& 
          0.737(7) & 12.339(7) & 11.987(7) & 
          0.755(7) & 12.357(7) & 12.005(7) \\
   6.5 &  0.670(10)&  6.958(10)&  6.770(10)& 
          0.739(6) &  7.027(6) &  6.839(6) & 
          0.758(5) &  7.046(5) &  6.858(5) \\
   4.5 &  0.665(8) &  5.021(8) &  4.892(8) & 
          0.738(5) &  5.093(5) &  4.965(5) & 
          0.758(4) &  5.113(4) &  4.985(4) \\
   3.8 &  0.663(8) &  4.341(8) &  4.235(8) & 
          0.737(5) &  4.416(5) &  4.309(5) & 
          0.757(4) &  4.436(4) &  4.330(4) \\
   3.0 &  0.658(7) &  3.564(7) &  3.481(7) & 
          0.735(4) &  3.642(4) &  3.559(4) & 
          0.756(4) &  3.663(4) &  3.580(4) \\
    \hline
    \multicolumn{10}{c}{$\beta$=5.9}\\ \hline
  10.0 &  0.531(8) & 10.244(8) & 10.058(8) & 
          0.580(5) & 10.292(5) & 10.106(5) & 
          0.591(5) & 10.303(5) & 10.117(5) \\
   5.0 &  0.528(7) &  5.389(7) &  5.298(7) & 
          0.575(4) &  5.435(4) &  5.345(4) & 
          0.586(4) &  5.446(4) &  5.356(4) \\
   3.0 &  0.522(6) &  3.440(6) &  3.387(6) & 
          0.569(3) &  3.488(3) &  3.435(3) & 
          0.580(3) &  3.498(3) &  3.446(3) \\
   2.1 &  0.511(6) &  2.601(6) &  2.594(6) & 
          0.560(3) &  2.650(3) &  2.643(3) & 
          0.571(3) &  2.661(3) &  2.654(3) \\
   1.3 &  0.487(5) &  1.806(5) &  1.818(5) & 
          0.539(3) &  1.859(3) &  1.871(3) & 
          0.552(2) &  1.871(2) &  1.883(2) \\
    \hline
    \multicolumn{10}{c}{$\beta$=6.1}\\ \hline
   7.0 &  0.435(8) &  7.255(8) &  7.158(8) & 
          0.471(5) &  7.292(5) &  7.195(5) & 
          0.479(5) &  7.299(5) &  7.202(5) \\
   3.5 &  0.423(7) &  3.835(7) &  3.788(7) & 
          0.462(4) &  3.875(4) &  3.828(4) & 
          0.470(4) &  3.883(4) &  3.836(4) \\
   2.1 &  0.408(6) &  2.499(6) &  2.494(6) & 
          0.450(3) &  2.541(3) &  2.535(3) & 
          0.458(3) &  2.549(3) &  2.544(3) \\
   1.5 &  0.393(5) &  1.903(5) &  1.908(5) & 
          0.436(3) &  1.946(3) &  1.951(3) & 
          0.445(3) &  1.954(3) &  1.959(3) \\
   0.9 &  0.359(4) &  1.320(4) &  1.352(4) & 
          0.404(3) &  1.364(3) &  1.396(3) & 
          0.413(2) &  1.373(3) &  1.405(2) \\
  \end{tabular}
  \caption{Binding energy and the total mass of the
    heavy-light pseudoscalar mesons.}
  \label{tab:ps_mass}
\end{table}

\begin{table}
  \begin{tabular}{c|c|c|c}
    $aM_{0}$ & $\kappa=\kappa_{\mbox{\scriptsize  crit}}$ 
             & $\kappa=\kappa_{s1}$
             & $\kappa=\kappa_{s2}$ \\   \hline
    \multicolumn{4}{c}{$\beta$=5.7}\\ \hline
$\infty$&
        0.675(41)& 0.814(34)& 0.851(36)\\
 12.0 & 0.588(25)& 0.693(19)& 0.722(20)\\
  6.5 & 0.531(19)& 0.615(13)& 0.638(13)\\
  4.5 & 0.481(15)& 0.556(11)& 0.575(10)\\
  3.8 & 0.456(14)& 0.527(9) & 0.546(9) \\
  3.0 & 0.421(12)& 0.486(8) & 0.503(8) \\
    \hline
    \multicolumn{4}{c}{$\beta=5.9$}\\ \hline
$\infty$&
        0.312(15)& 0.370(11)& 0.383(11)\\
 10.0 & 0.285(11)& 0.333(8) & 0.344(7) \\
  5.0 & 0.260(9) & 0.296(7) & 0.304(7) \\
  3.0 & 0.235(8) & 0.264(5) & 0.271(5) \\
  2.1 & 0.213(7) & 0.240(4) & 0.246(4) \\
  1.3 & 0.178(5) & 0.201(3) & 0.207(3) \\
    \hline
    \multicolumn{4}{c}{$\beta=6.1$}\\ \hline
$\infty$&
        0.178(12)& 0.205(9) & 0.210(8) \\
  7.0 & 0.159(9) & 0.185(7) & 0.190(6) \\
  3.5 & 0.140(7) & 0.165(5) & 0.170(4) \\
  2.1 & 0.124(5) & 0.148(4) & 0.152(3) \\
  1.5 & 0.114(4) & 0.135(3) & 0.140(3) \\
  0.9 & 0.096(3) & 0.114(2) & 0.118(2) \\
  \end{tabular}
  \caption{Raw data of $a^{3/2}(f_P\sqrt{M_P})^{(0)}$
    at $\kappa_{\mbox{\scriptsize crit}}$,
       $\kappa_{s1}$ and $\kappa_{s2}$.}
  \label{tab:fsqrtm_rawdata_f^0}
\end{table}

\begin{table}
  \begin{tabular}{c|c|c|c}
    $aM_{0}$ & $\kappa=\kappa_{\mbox{\scriptsize crit}}$
             & $\kappa=\kappa_{s1}$
             & $\kappa=\kappa_{s2}$ \\   \hline
    \multicolumn{4}{c}{$\beta$=5.7}\\ \hline
$\infty$&
        $-$0.485(34)& $-$0.556(27)& $-$0.576(29)\\
 12.0 & $-$0.455(22)& $-$0.511(16)& $-$0.526(16)\\
  6.5 & $-$0.436(18)& $-$0.482(12)& $-$0.495(12)\\
  4.5 & $-$0.415(15)& $-$0.458(10)& $-$0.470(10)\\
  3.8 & $-$0.403(14)& $-$0.446(9) & $-$0.458(9) \\
  3.0 & $-$0.387(12)& $-$0.429(8) & $-$0.441(8) \\
    \hline
    \multicolumn{4}{c}{$\beta=5.9$}\\ \hline
$\infty$&
        $-$0.194(12)& $-$0.226(8) & $-$0.234(8) \\
 10.0 & $-$0.189(8) & $-$0.215(6) & $-$0.221(5) \\
  5.0 & $-$0.183(7) & $-$0.203(5) & $-$0.208(5) \\
  3.0 & $-$0.176(7) & $-$0.193(4) & $-$0.198(4) \\
  2.1 & $-$0.170(6) & $-$0.187(4) & $-$0.191(4) \\
  1.3 & $-$0.158(5) & $-$0.176(3) & $-$0.180(3) \\
    \hline
    \multicolumn{4}{c}{$\beta=6.1$}\\ \hline
$\infty$&
        $-$0.098(8) & $-$0.111(6) & $-$0.113(5) \\
  7.0 & $-$0.092(6) & $-$0.105(4) & $-$0.108(4) \\
  3.5 & $-$0.086(5) & $-$0.100(3) & $-$0.103(3) \\
  2.1 & $-$0.082(4) & $-$0.097(3) & $-$0.100(3) \\
  1.5 & $-$0.080(4) & $-$0.095(2) & $-$0.098(2) \\
  0.9 & $-$0.079(3) & $-$0.093(2) & $-$0.096(2) \\
    \end{tabular}
  \caption{Raw data of 
    $2aM_0a^{3/2}(f_P\sqrt{M_P})^{(1)}$
    at $\kappa_{\mbox{\scriptsize crit}}$, $\kappa_{s1}$ and $\kappa_{s2}$.}
  \label{tab:fsqrtm_rawdata_f^1}
\end{table}

\clearpage
\narrowtext
\begin{table}
  \begin{tabular}{c|ccc}
                & $\beta=6.1$ & $\beta=5.9$ & $\beta=5.7$ \\ \hline
    tree        & 1.903(5) & 2.710(6) & 4.206(8) \\
    $q^*=\pi/a$ & 1.913(6) & 2.765(6) & 4.345(9) \\
    $q^*=1/a$   & 1.919(6) & 2.804(7) & 4.476(9) \\
  \end{tabular}
  \caption{Bare b-quark mass that reproduces the physical
    $B$ meson mass.
    }
  \label{tab:b-quark_mass}
\end{table}

\begin{table}
  \begin{tabular}{c|ccc|ccc|ccc}
    &\multicolumn{3}{c|}{$\beta=6.1$}
    &\multicolumn{3}{c|}{$\beta=5.9$}
    &\multicolumn{3}{c}{$\beta=5.7$}\\   \hline
    & tree & $q^*=\pi/a$ & $q^*=1/a$ &
      tree & $q^*=\pi/a$ & $q^*=1/a$ &
      tree & $q^*=\pi/a$ & $q^*=1/a$  \\ \hline
    $f_{B}$&
              0.184(7) & 0.166(7) & 0.157(6) &
              0.210(6) & 0.181(6) & 0.163(5) &
              0.233(7) & 0.189(6) & 0.150(5) \\
    $f_{B_{s}}$ ($\kappa_{s1}$) &
              0.215(5) & 0.195(5) & 0.185(4) &
              0.233(4) & 0.203(4) & 0.183(3) &
              0.265(5) & 0.217(4) & 0.174(3) \\
    $f_{B_{s}}$ ($\kappa_{s2}$) &
              0.222(5) & 0.201(4) & 0.190(4) &
              0.239(4) & 0.208(4) & 0.187(3) &
              0.274(5) & 0.225(4) & 0.180(3) \\
  \end{tabular}
  \caption{Results for $f_B$ and $f_{B_s}$ in GeV.
    }
  \label{tab:decay_constant_results}
\end{table}

\begin{table}[htbp]
  \begin{center}
    \begin{tabular}{c|ccc}
                & $\beta=6.1$ & $\beta=5.9$ & $\beta=5.7$ \\
       \hline 
       $O(\alpha_s^2)$         & 3\% & 5\% & 12\% \\
       $O((a\Lambda_{QCD})^2)$ & 2\% & 3\% &  8\% \\
       $O(\alpha_s/(aM)^2)$    & 6\% & 4\% &  2\% \\
    \end{tabular}
    \caption{An order estimate of the possible systematic
       errors. 
}
    \label{tab:systematic_errors}
  \end{center}
\end{table}

\end{document}